%% file: samplepaper.tex
\documentclass[runningheads]{llncs}
\usepackage[T1]{fontenc}
\usepackage{graphicx}
\input{sections/commands}
\begin{document}
\title{From Sands to Mansions: Actionable, Customizable and Causality-Preserving Cyberattack Emulation with LLM-powered Symbolic Planning}
\titlerunning{Cyberattack Emulation with LLM-powered Symbolic Planning}
\author{Lingzhi Wang\inst{1} \and
Zhenyuan LI\inst{2} \and
Yi Jiang\inst{2} \and
Zhengkai Wang\inst{2} \and
Xiangmin Shen\inst{3} \and
Wei Ruan\inst{2} \and
Yan Chen\inst{1}
}
\authorrunning{L. Wang et al.}
\institute{Northwestern University, Evanston, IL 60208, USA \\
\email{lingzhiwang2025@u.northwestern.edu} \\
\email{ychen@northwestern.edu} \and
Zhejiang University, Hangzhou, Zhejiang 310027, China \\
\email{\{lizhenyuan,22421062,22451237,ruanwei\}@zju.edu.cn} \and
Hofstra University, Hempstead, NY 11549, USA\\
\email{xiangmin.shen@hofstra.edu}}
\maketitle              % typeset the header of the contribution
\begin{abstract}
\input{sections/Abstract}

\end{abstract}
\section{Introduction}
\label{sec:intro}
\input{sections/Introduction}

\section{Background and Related Work}
\label{sec:background}
\input{sections/Background}

% \input{sections/ThreatModel}

% \subsection{Motivating Examples}
% \label{sec:examples}
% \input{sections/Examples}

% \section{Problem Definition and Assumptions}
% \label{sec:problem_definition}
% \input{sections/Definition}

\section{Cyberattack Planning}
\label{sec:linking}
\input{sections/Linking_Model}

\section{System Design}
\label{sec:design}
\input{sections/Design}

% \section{Application}
% \label{sec:application}
% \input{sections/Application}

% \section{Implementation}
% \label{sec:implementation}
% \input{sections/Implementation}

\section{Evaluation}
\label{sec:evaluation}
\input{sections/Evaluation}

% \section{Related Work}
% \label{sec:related}
% \input{sections/Related_Work}

\section{Limitation and Future Work}
\label{sec:discussion}
\input{sections/Discussion}

\section{Conclusion}
\label{sec:conclusion}
\input{sections/Conclusion}

\section{Ethics Considerations}
\label{sec:ethical}
\input{sections/Ethical}

\begin{credits}
\subsubsection{\ackname} We would like to thank anonymous reviewers for their constructive feedback.
This project was supported by the National Science Foundation (NSF) grant 2148177 and funds from the Resilient \& Intelligent NextG Systems (RINGS) program. 

\subsubsection{\discintname}
The authors declare that they have no competing interests.
\end{credits}
\bibliographystyle{splncs04}
\bibliography{refs}




% LNCS Homepage, \url{http://www.springer.com/lncs}, last accessed 2023/10/25
% \end{thebibliography}

\appendix
\input{sections/Appendix}

\end{document}

%% file: sections/commands.tex
\newcommand{\SysName}{{\scshape Aurora}\xspace}
\usepackage{etoolbox}
\usepackage{xspace}
\usepackage{pifont}
\usepackage{makecell}
\newcommand{\NODOZE}{{\scshape NoDoze}}

\newcommand{\PROVDETECTOR}{{\scshape ProvDetector}}

\newcommand{\FLASH}{{\scshape Flash}}

\newcommand{\LADDER}{{\scshape LADDER}}

\usepackage{listings}
\usepackage{xcolor}
\usepackage{graphicx}
\usepackage{tabularx}
\usepackage{makecell}
\usepackage{algorithm}
\usepackage{booktabs}
\usepackage{multirow}
\usepackage{algpseudocode}
\usepackage{dcolumn}
\usepackage{enumitem}
\usepackage{hyperref}

\usepackage{array}

\newcolumntype{d}[1]{D{.}{.}{#1}}

\colorlet{punct}{red!60!black}
\definecolor{background}{HTML}{EEEEEE}
\definecolor{delim}{RGB}{20,105,176}
\colorlet{numb}{magenta!60!black}

\lstdefinelanguage{rules}{
    basicstyle=\footnotesize\ttfamily,
    numberstyle=\scriptsize,
    stepnumber=1,
    numbersep=8pt,
    showstringspaces=false,
    breaklines=true,
    frame=single,
    frameround = tttt,
    captionpos = b,
    backgroundcolor=\color{background},
    literate=
     *{-}{{{\color{delim}{-}}}}{1}
     {name:}{{{\color{punct}{name:}}}}{5}
}

\lstdefinelanguage{yml}{
    basicstyle=\footnotesize\ttfamily,
    numberstyle=\scriptsize,
    stepnumber=1,
    numbersep=8pt,
    showstringspaces=false,
    breaklines=true,
    frame=single,
    frameround = tttt,
    captionpos = b,
    backgroundcolor=\color{background},
    literate=
     *{-}{{{\color{delim}{-}}}}{1}
     {name:}{{{\color{punct}{name:}}}}{5}
}

\lstdefinelanguage{json}{
    basicstyle=\footnotesize\ttfamily,
    numbers=left,
    numberstyle=\scriptsize,
    stepnumber=1,
    numbersep=8pt,
    showstringspaces=false,
    breaklines=true,
    frame=single,
    frameround = tttt,
    captionpos = b,
    backgroundcolor=\color{background},
    literate=
     *{0}{{{\color{numb}0}}}{1}
      {1}{{{\color{numb}1}}}{1}
      {2}{{{\color{numb}2}}}{1}
      {3}{{{\color{numb}3}}}{1}
      {4}{{{\color{numb}4}}}{1}
      {5}{{{\color{numb}5}}}{1}
      {6}{{{\color{numb}6}}}{1}
      {7}{{{\color{numb}7}}}{1}
      {8}{{{\color{numb}8}}}{1}
      {9}{{{\color{numb}9}}}{1}
      {:}{{{\color{punct}{:}}}}{1}
      {,}{{{\color{punct}{,}}}}{1}
      {\{}{{{\color{delim}{\{}}}}{1}
      {\}}{{{\color{delim}{\}}}}}{1}
      {[}{{{\color{delim}{[}}}}{1}
      {]}{{{\color{delim}{]}}}}{1},
}

\lstdefinelanguage{prompt}{
    basicstyle=\footnotesize\ttfamily,
    numbers=left,
    numberstyle=\scriptsize,
    stepnumber=1,
    numbersep=8pt,
    showstringspaces=false,
    breaklines=true,
    frame=single,
    frameround = tttt,
    captionpos = b,
    backgroundcolor=\color{background},
    moredelim=[is][\textcolor{punct}]{@}{@}, % highlight
    literate=
     *{0}{{{\color{numb}0}}}{1}
      {1}{{{\color{numb}1}}}{1}
      {2}{{{\color{numb}2}}}{1}
      {3}{{{\color{numb}3}}}{1}
      {4}{{{\color{numb}4}}}{1}
      {5}{{{\color{numb}5}}}{1}
      {6}{{{\color{numb}6}}}{1}
      {7}{{{\color{numb}7}}}{1}
      {8}{{{\color{numb}8}}}{1}
      {9}{{{\color{numb}9}}}{1}
      {\{}{{{\color{delim}{\{}}}}{1}
      {\}}{{{\color{delim}{\}}}}}{1}
      {[}{{{\color{delim}{[}}}}{1}
      {]}{{{\color{delim}{]}}}}{1},
}

\lstdefinelanguage{report}{
    basicstyle=\footnotesize\ttfamily,
    numberstyle=\scriptsize,
    stepnumber=1,
    numbersep=8pt,
    showstringspaces=false,
    breaklines=true,
    breakindent=0pt,
    frame=single,
    frameround = tttt,
    captionpos = b,
    backgroundcolor=\color{background},
    xleftmargin=0pt,
}

%% file: sections/Abstract.tex
Evolving attacker capabilities demand realistic and continuously updated cyberattack emulation for threat-informed defense and security benchmarking.
Towards automated attack emulation, this paper defines modular attack actions and a linking model to organize and chain heterogeneous attack tools into causality-preserving cyberattacks.
Building on this foundation, we introduce \SysName: an automated cyberattack emulation system powered by symbolic planning and large language models (LLMs).
\SysName crafts actionable, causality-preserving attack chains tailored to Cyber Threat Intelligence (CTI) reports and target environments, and automatically executes these emulations.
Using \SysName, we generated an extensive cyberattack emulation dataset from 250 attack reports, 15 times larger than the leading expert-crafted dataset.
Our evaluation shows that \SysName significantly outperforms existing methods in creating actionable, diverse, and realistic attack chains.
We release the dataset and use it to evaluate three state-of-the-art intrusion detection systems, whose performance differed notably from results on older datasets, highlighting the need for up-to-date, automated attack emulation.

\textbf{This is the preprint version of a paper accepted to the 24th Applied Cryptography and Network Security (ACNS 2026).}

%% file: sections/Introduction.tex
The continuous evolution of cyberattack technologies makes comprehensive and up-to-date attack emulation essential for benchmarking defense systems, identifying their weaknesses, and enhancing threat-informed defenses. 
Consequently, attack emulation has become increasingly valued by governments, organizations, and security vendors.
Notable examples include the European Central Bank's threat intelligence-based ethical red teaming (TIBER-EU) framework~\cite{tibereu}, purple team exercises conducted by the U.S. Cybersecurity and Infrastructure Security Agency and the National Security Agency, and numerous commercial products~\cite{cymulate,pentera,horizon3} in this area.
Since 2018, MITRE has annually organized expert emulation of advanced persistent threat (APT) attacks, attracting participation from over 50 top security vendors worldwide.
The growing importance of this field is also underscored by market projections, with MarketsandMarkets~\cite{marketingreport} forecasting the automated cyberattack emulation market to expand from \$729.2 million in 2024 to over \$2.4 billion by 2029.

Despite growing interest, there remains a notable scarcity of \textit{actionable}, \textit{causally coherent}, \textit{scalable}, and \textit{customizable} cyberattack emulation systems, especially for multi-stage APT attacks~\cite{takahashi2020aptgen,choi2021probabilistic}.
% An ideal cyberattack emulation tool should be Actionable, Diverse, Realistic, and Customizable.
An actionable attack emulation system should provide detailed commands, tools, and instructions to reproduce emulated attacks.
This facilitates multi-layer data collection (e.g., network traffic, authentication events, and host-based system logs) according to specific needs.
Second, effective attack emulation must preserve causal relationships between attack steps, ensuring that each step is executed based on the state and artifacts produced by previous steps, thereby accurately reflecting real-world attacker behavior.
Furthermore, as new attack techniques emerge daily, a scalable emulation system must easily incorporate new techniques without architectural changes and automatically generate diverse attack chains.
% Given the significant human expertise and effort required in attack emulation, we aim to automatically leverage existing threat intelligence to enhance the diversity.
Finally, we consider an emulation system to be customizable if it can automatically generate attack chains that are tailored to a specific target environment or aligned with the behaviors of a particular threat actor, rather than relying on manually authored playbooks or scripts.

\newcommand{\cmark}{\ding{51}}
\newcommand{\xmark}{\ding{55}}
\begin{table*}[t]
    \centering
    \scriptsize
    
    \caption{Comparison of existing efforts in attack emulation. The checkmarks indicate whether a system satisfies each property under the definitions used in this paper. Partial or manual support is not considered sufficient.}
    \label{tab:baseline_comparison}
% \begin{tabular}{l|l||c|c|c|c|c|c}
% \toprule
% \multirow{2}{*}{Category}   & \multirow{2}{*}{Examples}   & \multirow{2}{*}{Actionable} & \multicolumn{3}{c|}{Realism} & \multirow{2}{*}{Expert-Free} & \multirow{2}{*}{Scalable} \\
% & & & Multi-Step & Connectivity & Report Alignment &  &  \\
% \midrule
% Auditing Trace Dataset     & DARPA TC/OpTC                      & \xmark          & \cmark    & \cmark    & Unknown   & \xmark           & \xmark        \\
% Attack Graph Generation        & \multicolumn{1}{l}{}           & \xmark          & \cmark    & \xmark      & \cmark       & -           & -        \\
% Single-point Testing Tools & ART, Metasploit, Tools             & \cmark        & \xmark      & -      & \xmark         & -           & -        \\
% Attack Emulation Tools     & PurpleSharp, Attack Range, Caldera & \cmark        & \cmark    & \xmark      & \xmark         & -           & -        \\
% Human-Crafted Attacks      & MITRE Evaluation                   & \cmark        & \cmark    & \cmark    & \cmark       & \xmark           & \xmark        \\
% \midrule
% \ZY{?}                     & \ZY{\SysName}                & \cmark        & \cmark    & \cmark    & \cmark       & \cmark         & \cmark   \\  
% \bottomrule
% \end{tabular}
\begin{tabular}{l|l||c|c|c|c}
\toprule
Category & Examples & Actionable & \makecell{Causally\\Coherent} & Customizable & Scalable\\
\midrule
\makecell{Human-Orchestrated \\ Attack Campaigns}   & DARPA TC, OpTC                       & \xmark & \cmark & \xmark  & \xmark \\ \hline
\makecell{Abstract Attack \\ Modeling}  & AttacKG                & \xmark & \cmark & \xmark   & \cmark      \\ \hline
\makecell{Single-point \\ Testing Tools} & Atomic Red Team              & \cmark & \xmark & \cmark   & \cmark      \\ \hline
\makecell{Procedural Attack \\Orchestration Tools}     & PurpleSharp, Caldera  & \cmark & \xmark & \xmark   & \cmark      \\ \hline
\makecell{Human-Crafted \\ Attack Playbooks}      & MITRE Evaluation                    & \cmark & \cmark & \xmark   & \xmark \\
\midrule
               \multicolumn{2}{c||}{\textbf{\SysName}}                     & \cmark & \cmark & \cmark   & \cmark \\
\bottomrule
\end{tabular}
\end{table*}

\begin{figure}[h]
    \centering
    \includegraphics[width=0.75\linewidth]{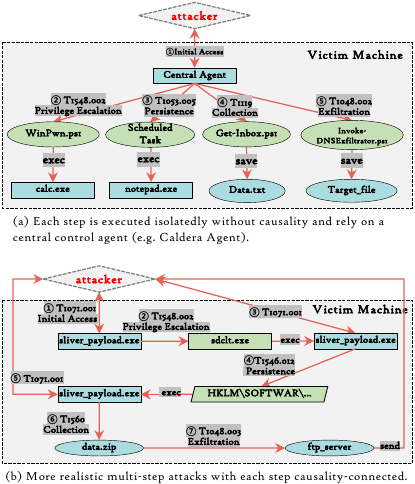}
    \caption{Causality in emulated attack chains}
    \label{fig:motivating_example}
    \vspace{-0.25in}
\end{figure}

Table~\ref{tab:baseline_comparison} highlights the limitations of existing efforts in emulating actionable, causally coherent, scalable, and customizable cyberattacks.
For instance, several widely used datasets~\cite{darpa_engagement,streamspot,alsaheel2021atlas} are derived from human-orchestrated red team campaigns and consist of post-incident audit logs.
While these attacks are realistic and causally coherent, they are recorded only as static traces.
As a result, such non-actionable datasets do not support replaying attacks for actively testing defense systems.
Single-point attack emulation tools~\cite{atomicredteam,holm2016sved,metasploit,viper} execute isolated tests and lack mechanisms to generate multi-stage, causally coherent attack chains.
While some procedural attack orchestration tools~\cite{caldera,purplesharp,attackrange} attempt to assemble multi-step attacks from isolated actions, they do not explicitly model inter-step dependencies, and therefore struggle to chain attacks comparable to real-world attack campaigns (see Figure~\ref{fig:motivating_example}).
Expert-crafted attack playbooks from MITRE Engenuity ATT\&CK Evaluations~\cite{citd_eval} offer high-fidelity, causally coherent attack representations, but their reliance on manual expertise limits their adaptability and scalability.

% Traditional vulnerability assessments and penetration testing merely focus on vulnerability exploitation and gaining system access without considering the post-exploitation steps.
% Red teaming, Breach and Attack Simulation (BAS)~\cite{bas_intro_1}, and Continuous Automated Red Teaming (CART), while capable of constructing more complete attack chains, primarily focus on generating a successful attack path for a specific target, thereby limiting the diversity of the attacks produced.
% Moreover, all these approaches heavily depend on skilled security experts.
% According to our discussions with some commercial BAS and CART providers, most of these systems are proprietary and come with high acquisition prices.

Three core challenges hinder the development of ideal cyberattack emulation systems.
The first challenge lies in the unstructured nature of cyberattack knowledge, which requires significant human expertise to integrate diverse Tactics, Techniques, and Procedures (TTPs) from heterogeneous sources into coherent attack chains.
Consequently, many existing approaches cover only a small subset of TTPs; for instance, the state-of-the-art automated attack planner ChainReactor~\cite{depasquale_chainreactor} primarily focuses on privilege escalation and considers merely about 30 attack actions.
The second challenge lies in constructing multi-step and causality-preserving attack chains, which requires understanding the relationship between different attack steps.
Existing studies~\cite{applebaum2016intelligent,depasquale_chainreactor,hoffmann2015simulated} have yet to propose a formal framework for modeling these relationships.
The third challenge is tailoring attacks to specific environments and threat intelligence, which ensures emulations both reflect real-world adversary behavior from CTI reports while remaining compatible with the tools and constraints of the target environment.
To our knowledge, transforming CTI into attacks that can actually be executed in real environments remains an open problem.

To address these challenges, we propose \SysName, the first automated, causality-preserving, customizable cyberattack emulation system.
\SysName converts third-party attack tools into standard, modular attack actions using LLM.
These actions are then interconnected into causality-preserved attack chains with symbolic planning.
\SysName can customize attack chains based on the emulation environment, available tools, and intelligence derived from CTI reports.
It also provides actionable Python scripts to execute these attacks automatically.
Experimental results demonstrate that \SysName's superior capability in building actionable, diverse, and causality-preserved attacks compared to existing approaches.
\SysName has extracted over 5,500 attack actions from five popular third-party attack tools.
Leveraging these actions, we constructed attacks based on 250 CTI reports.
These attacks are published as a continually growing cyberattack dataset for future research.

In this paper, we make the following contributions.
First, we define attack actions to modularize heterogeneous attack tools and propose a novel attack action linking model to connect them into coherent attack chains.
Second, we introduce \SysName, the first automated, customizable, and causality-preserving cyberattack emulation system.
Third, experiments show that \SysName generates more causality-preserved attack chains than other automated approaches and offers greater scalability and customization than expert-crafted attacks.
We use the generated attacks to evaluate three state-of-the-art intrusion detection systems, whose performance differed notably from results on older datasets.
To facilitate future research, we open-sourced the attack dataset generated by \SysName\footnote{\url{https://github.com/LexusWang/Aurora-demos}},
which consists of over 250 unique attack chains, 15 times larger than the leading expert-crafted equivalent.

%% file: sections/Background.tex
\subsection{Cyberattack Emulation}\label{sec:background-emulation}
\begin{figure}[tb]
    \centering
    \includegraphics[width=0.55\linewidth]{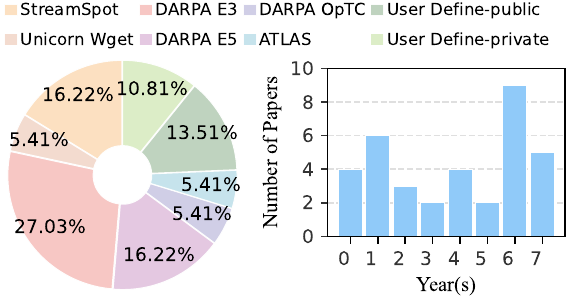}
    \caption{Benchmark datasets in intrusion detection research (Left: Proportional usage of datasets. Right: Dataset age at the time of paper publication.)}
    \label{fig:dataset_review}
    \vspace{-0.2in}
\end{figure}

The need for open, comprehensive, and continuously updated cyberattack emulation is widely recognized by both industry and the research community~\cite{choi2021probabilistic,takahashi2020aptgen}.
We reviewed representative prior efforts in cyberattack emulation, including papers published in major security venues, as well as widely adopted open-source tools in the industry.
Specifically, we survey (1) benchmark datasets commonly adopted in recent intrusion detection papers,
(2) abstract attack modeling approaches focusing on high-level attack representation,
(3) executable public cyberattack testing and emulation tools, and
(4) expert-curated attack emulations used as industry benchmarks.
However, no existing work satisfies our needs.
This gap is reflected by the limited and outdated cyberattack benchmark datasets in recent intrusion detection papers~\cite{rehman2024flash,li2023nodlink,cheng2024kairos,jia2024magic,goyal2024r,wang2024incorporating,yang2023prographer,zengy2022shadewatcher,milajerdi2019poirot,morse,han2020unicorn,holmes,nodoze,wang2020you,jian2025}.
Our analysis (shown in Figure~\ref{fig:dataset_review}) reveals two critical problems.
First, the benchmarking is dominated by a few non-scalable datasets~\cite{darpa_engagement,streamspot,alsaheel2021atlas}, indicating a narrow and repetitive evaluation scope.
Second, detection systems are often evaluated against outdated threats.
% Moreover, these attack datasets are not scalable since they are produced by professional red teams or labs manually. 

Many researches~\cite{choi2021probabilistic,takahashi2020aptgen,ttpdrill,extractor,li2022attackg,alam2023looking,rahman2023attackers,kumarasinghe2024semantic} focus on abstract attack modeling, aiming to capture attacker behaviors at the semantic level using probabilistic models or CTI reports.
Their primary goal is to model \textit{what} attackers do, typically in terms of MITRE ATT\&CK TTPs, to support threat understanding, detection, and attribution.
However, such models remain inherently abstract.
They operate at the level of TTP sequences and do not provide actionable specifications, such as concrete tools, commands, scripts, execution contexts, or inter-step state dependencies.
As a result, they cannot be directly instantiated or replayed as executable attacks in real-world environments.

To address the lack of actionability in abstract attack models, many systems provide executable implementations of attack techniques.
These systems can execute attack actions in controlled environments.
However, most of these tools (e.g., Atomic Red Team~\cite{atomicredteam}, Metasploit~\cite{metasploit}, Viper~\cite{viper}, and LOLBAS~\cite{LOLBAS}) execute individual techniques or actions in isolation rather than as a coherent attack campaign.
Executing isolated steps is unable to emulate real-world multi-step attacks, and thus insufficient for evaluating defense systems that rely on cross-step correlation and contextual reasoning~\cite{xu2024autoattacker,holmes,cheng2023kairos}.
For instance, file compression is typically benign when performed in isolation, but becomes suspicious when preceded by data staging and followed by exfiltration.

To move beyond isolated testing, several attack emulation frameworks aim to construct multi-step attacks by orchestrating atomic actions into predefined playbooks.
Representative examples include Caldera~\cite{caldera}, PurpleSharp~\cite{purplesharp}, and Attack Range~\cite{attackrange}.
% These systems represent an important step toward automation by enabling the execution of multiple attack actions in sequence.
However, the resulting attack chains remain largely procedural.
The actions are executed sequentially without explicitly modeling the causal relationships or shared objectives that connect them into a coherent attack campaign.
As a result, the emulated attacks often deviate from real-world attacker behavior (see Figure~\ref{fig:motivating_example}).
For example, escalating privileges solely to execute \texttt{calc.exe} serves no meaningful attack.
We refer to such systems as \textit{procedural attack orchestration frameworks}, as they focus on sequencing predefined actions rather than composing attacks based on causal dependencies.
Moreover, the playbooks used by these frameworks are manually authored, and there is no mechanism for automatically generating or adapting playbooks based on different environments or attacker profiles.
This reliance on human authoring fundamentally limits both the scalability and the customization of attack emulation.

Given those shortcomings, human-crafted attacks are still a reliable source for attack emulation~\cite{citd_eval,apt_attack_simulation}.
MITRE ATT\&CK Evaluations~\cite{citd_eval} provide one of the most well-recognized and public attack emulations.
These attacks are fully actionable, logically coherent, and aligned with CTI reports.
However, their generation requires immense manual effort and deep domain expertise.
This is evident in the small number of scenarios produced (e.g., 17 attack chains from 15 adversary groups over eight years) and the substantial fees required for vendor participation.
% Consequently, this manual approach is fundamentally unscalable and cannot produce the large, diverse datasets needed by the research community.
Table~\ref{tab:baseline_comparison} shows the comparison of existing work.

\vspace{-0.1in}
\subsection{Symbolic Planning}\label{sec:background_pddl}
In general, a planning problem aims to find a sequence of actions that achieves a specific goal~\cite{chen2024language,ding2023integrating}.
Specifically, a symbolic planning problem can be defined as a tuple $(\mathcal{P}, \mathcal{A}, I, G)$, where $\mathcal{P}$ is a finite set of predicates and $\mathcal{A}$ is a finite set of actions.
A predicate ($p \in \mathcal{P}$) is a boolean proposition describing a specific condition of the target world, and a subset of $\mathcal{P}$ can represent a state $s$ of the target world at some stage.
$I$ and $G$ are two special states: $I$ is the initial state, and $G$ is the goal state.
An action ($a \in \mathcal{A}$) is the basic unit that can change the states.
It includes two features: the \textit{preconditions}, which are the predicates that must be true before the action can be applied, and the \textit{effects}, which are the predicates that will be true after the action is applied.
A plan is a sequence of actions changing the target world from the initial state to the goal state.
Symbolic planning typically employs \textit{Planning Domain Definition Language} (PDDL), a declarative language, to define planning problems with symbolic notations.

Cyberattacks are often modeled as symbolic planning problems.
However, existing work either lacks clear definitions of actions and states~\cite{phillips1998graph,hu2020automated,loevenich2025agentic,loevenich2025automating}, or relies on overly abstract or limited action spaces~\cite {sarraute2012pomdps,hoffmann2015simulated,miller2018automated,li2024dynpen,holm2022lore,applebaum2016intelligent} that fail to reflect real-world complexity~\cite{mitre_matrix}.
More importantly, most existing approaches do not explicitly model how actions are causally linked through preconditions and effects.
Some approaches~\cite{loevenich2025agentic,loevenich2025automating} do not specify how one-shot planning can be applied to generate multi-stage APT campaigns, where each stage has distinct initial and goal states.
The absence of a well-defined action space, a principled linking model, and a multi-stage planning schema leaves a critical gap in planning actionable and causally coherent cyberattacks.

\subsection{Assumptions and Scope}
Unlike penetration testing and red teaming, which aim to discover vulnerabilities or attack paths in unknown or partially known systems, the goal of attack emulation is to construct and execute emulated attack scenarios within a known environment.
Accordingly, \SysName assumes that basic information about the target testbed, including operating system versions, installed software, known vulnerabilities, and network topology, is available for emulation.
Such information can be automatically collected using existing asset inventory and security assessment tools (e.g., OpenSCAP, OpenVAS, and osquery~\cite{openscap,openvas,osquery}).
In this paper, we do not consider planning under incomplete information.
Importantly, this assumption does not restrict \SysName to a single predefined testbed or static environment, as in some prior work~\cite{citd_eval,attackrange}. Instead, \SysName supports customized attack emulation across different user environments, as long as basic system information is available (see \S\ref{sec:starting-state} for details).

%% file: sections/Linking_Model.tex
This section introduces two core concepts, \textit{Attack Actions} and \textit{Attack Action Linking Model (AALM)}.
Attack actions provide modular and structured representations of actionable steps in cyberattacks.
The attack action linking model defines the relationships between these actions and connects them into causality-preserved attack chains.

\subsection{Attack Actions}\label{sec:attackaction}
In symbolic planning, an action is the atomic unit in a plan.
Similarly, we define an attack action as an atomic operation in cyberattack emulations.
\begin{definition}[Attack Action]
    An attack action is the smallest unit that is considered to form an attack plan and an atomic operation to execute in a multi-step attack.
\end{definition}

The granularity of an attack action depends on how attack tools are invoked in practice.
For example, an individual test in Atomic Red Team~\cite{atomicredteam}, a module in Metasploit~\cite{metasploit}, or a single command in post-exploitation frameworks~\cite{sliver,cobaltstrike} can each be treated as an attack action.
As summarized in Table~\ref{tab:basic_action_features}, each attack action has several associated attributes.
\textit{Name}, \textit{source}, and \textit{description} define the basic information of the action.
% The \textit{supported platform} field specifies the operating systems on which the action can be executed.
\textit{Tactics and techniques} map the action to the entries in the MITRE ATT\&CK matrix.
The \textit{execution} field specifies both concrete execution instructions (e.g., commands or scripts) and the required executor, which denotes the execution context or agent.
For example, a malicious PowerShell script uses \texttt{PowerShell} as its executor, whereas commands provided by Meterpreter~\cite{meterpreter} require an active Meterpreter session on the target system and therefore designate \texttt{Meterpreter} as their executor.
% Some actions involving user interaction (e.g., clicking a phishing link to download and execute a payload) specify ``User'' as the executor.
The \textit{precondition} and \textit{effect} attributes are essential for linking actions together.
Therefore, they are discussed in detail in the next section on the attack action linking model.

\begin{table}[ht!]
    \centering
    \scriptsize
    \caption{Attributes of attack actions}
    \begin{tabular}{ll}
    \toprule
    Field       & Description                         \\ \midrule
    UUID        & A universally unique identifier.    \\
    Name        & A short name of the attack action.        \\
    Description & A brief description of the attack action. \\
    Source      & The source tool of the attack action.   \\
    % Platform    & The operating systems supporting the action.   \\
    Tactics     & MITRE ATT\&CK tactics.              \\
    Technique   & MITRE ATT\&CK technique.            \\  
    Execution    & The concrete executor and instructions for execution. \\
    Preconditions     & Conditions that must be satisfied for action execution.              \\
    Effects   & Conditions that will be satisfied after action execution.             \\ \bottomrule  
    \end{tabular}
    \label{tab:basic_action_features}
\end{table}

The concept of attack actions is related to, but distinct from, MITRE ATT\&CK procedures~\cite{mitre_ttp}.
While procedures provide descriptions of how a technique may be implemented in practice, they are primarily intended for knowledge sharing and threat intelligence documentation.
As a result, procedures are expressed in unstructured natural language and do not explicitly define execution context, preconditions, or effects.
In contrast, attack actions are designed as executable and plannable primitives.
Each attack action explicitly specifies its executor, concrete instructions, preconditions, and effects, which enables attack actions to be automatically composed into causally coherent, actionable attack chains.
For example, a MITRE procedure may state that attackers dump credentials from LSASS using tools such as Mimikatz.
However, this description does not explicitly encode the prerequisite of administrative privileges or the resulting availability of credential artifacts.
In our model, credential dumping is represented as an attack action whose execution requires administrator access and whose effect is the acquisition of credentials.

% Therefore, we use a straightforward LLM approach to summarize and generate these names, without optimizing or evaluating accuracy in this paper.
% In cases where the documentation does not assign a specific name to each command, we use an LLM to generate a short name based on the content of the command documentation.
% This name helps human operators quickly understand each action, but it is not used in subsequent attack simulations.

% \begin{definition}[Attack Action Linking Model]
%     The attack action linking model is an extensible set of PDDL predicates, which is used to describe the preconditions and effects of an attack action.
% \end{definition}

% \begin{figure}[h!]
%     \centering
%     \includegraphics[width = \linewidth]{figures/attack_chain_example-pddl.pdf}
%     \caption{An example attack chain connected by predicates.}
%     \label{lst:attack_chain_example}
% \end{figure}

\vspace{-0.15in}
\subsection{Attack Action Linking Model}\label{sec:aalm}
As introduced in \S\ref{sec:background_pddl}, preconditions and effects are used to connect actions in symbolic planning.
Preconditions represent the requirements that must be met before executing an attack action.
For example, executing Sliver commands requires an active session on the victim machine, and data exfiltration requires the data to be collected.
Effects are the resulting conditions after executing an attack action.
An action is included in an attack chain only if 1) all its preconditions are fulfilled, and 2) its effects help satisfy the preconditions of the following actions or achieve specific attack goals.
However, the term \textit{condition} is broad, subjective, and ambiguous.
In this paper, we propose the Attack Action Linking Model (AALM), a standard, universal, and succinct framework, to describe the preconditions and effects from the perspective of cyberattack emulation.

\begin{definition}[Attack Action Linking Model]
    The attack action linking model is an extensible set of PDDL predicates used to describe the preconditions and effects of an attack action.
\end{definition}

We design the predicate set of AALM from nine dimensions: environments, executors, payloads, files, processes, users, information, data, and techniques.
Due to space limits, the full list is available in our open-source repository.

\begin{itemize}[left=3pt,]
    \item \textit{Environment} includes the predicates related to the operating systems, vulnerabilities, software, and network topology of the attack target.
    To standardize the predicate format across numerous vulnerabilities and software types, we use Common Vulnerabilities and Exposures (CVE)~\cite{cve} to identify vulnerabilities and Common Platform Enumeration (CPE)~\cite{cpe} to represent software.
    \item \textit{Executor} includes predicates about the executor type and states.
    Executor type encompasses common executors derived from popular attack frameworks and C2 tools~\cite{metasploit,atomicredteam,LOLBAS}.
    Executor states contains predicates about privilege levels, persistence status, and so on.
    \item \textit{Payload} includes predicates describing payload types, formats, and operations.
    The payload types and formats are derived from commonly used payload generation tools~\cite{msfvenom}.
    Payload operation predicates cover the conditions about loading, executing, and setting handlers.
    \item \textit{File} includes predicates related to file types, states, and operations.
    To ensure AALM covers the file types commonly involved in cyberattacks, we crawled documentation of popular open-source attack tools~\cite{atomicredteam,metasploit,LOLBAS} and cybersecurity websites~\cite{viper,mitre_eval} to extract a representative set of file types.
    It also covers similar predicates for directories and Windows registry entries.
    \item \textit{Information} includes predicates representing information involved in cyberattacks.
    Given the diversity of information types in cyberattacks, we organize them into 39 categories based on the technique types defined under the MITRE discovery tactic.
    \item \textit{Data} includes the predicates representing data types and data states.
    We categorize ten data types according to the technique taxonomy of the MITRE collection tactics.
    Data state predicates include the storage, transmission, and relay of data during an attack.
    \item \textit{User} includes the predicates representing user states, types, username, credentials, and privilege levels of users.
    \item \textit{Process} includes the predicates representing process states and process information such as process ID, name, and privilege level.
    \item \textit{Technique} includes the predicates that reflect the intent or consequence of an action.
    For example, although deleting backup files and deleting log files both result in \textsf{(file-deleted)}, their purposes are different.
    The former aims to inhibit system recovery, while the latter serves to remove indicators of compromise.
    Therefore, we design technique-level predicates such as \textsf{(inhibit-system-recovery)} and \textsf{(indicator-removal)} to show the difference of the effects.
\end{itemize}

\begin{figure}[ht!]
    \centering
    \includegraphics[width = 0.85\linewidth]{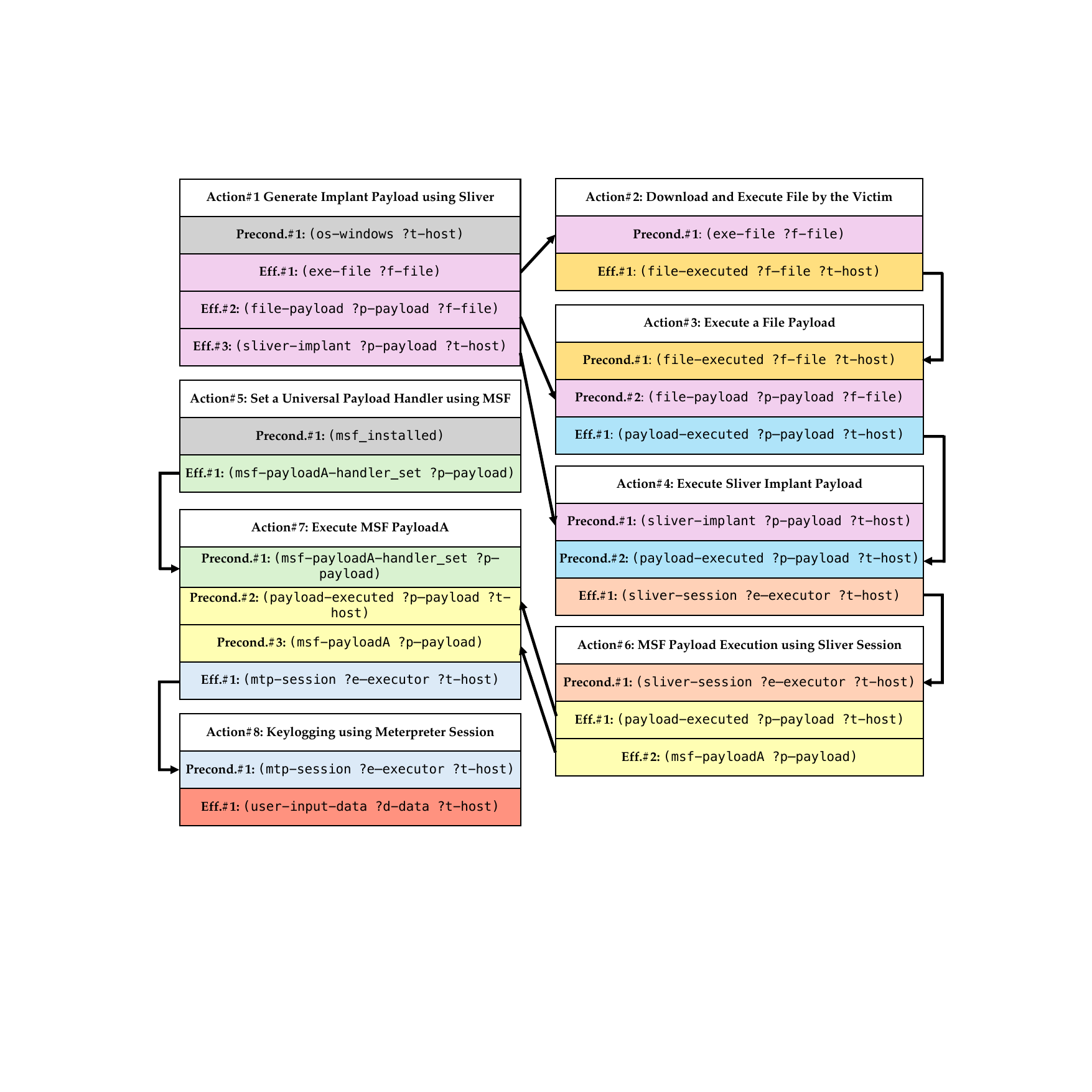}
    \caption{An attack chain example connected by AALM.}
    \label{lst:attack_chain_example}
\end{figure}

Figure~\ref{lst:attack_chain_example} shows an attack chain connected by predicates in AALM.
Given the complexity and evolving nature of cyberattacks, AALM is built with extensibility.
New predicates and actions can be incrementally added, similar to the ongoing expansion of the MITRE ATT\&CK framework.
AALM allows users to define custom actions along with their preconditions and effects using PDDL syntax.
AALM follows a monotonic extension principle: newly introduced predicates are incorporated through additional action preconditions and effects.
Existing attack chains remain unchanged and valid under the extended model.
For example, a user can introduce a predicate \texttt{(edr-service-disabled ?h)} as the effect of an evasion action and as the precondition for high-risk payload execution, explicitly modeling the causal relationship between disabling a specific endpoint protection and enabling stealthy execution.
As another example, when certain attack actions are only feasible with specific tools, users can introduce tool-specific predicates (e.g., a dedicated executor-type predicate) as the preconditions of those specialized actions.
% Such tool-specific predicates enable modeling exclusive relationships without affecting existing predicates and attack chains.
As the first work to propose a standardized, symbolic linking model for automated attack emulation, we envision its coverage and expressiveness to grow through collaboration with the broader security community.

%% file: sections/Design.tex
% \subsection{Overview}

\begin{figure}
    \centering
    \includegraphics[width=0.6\linewidth]{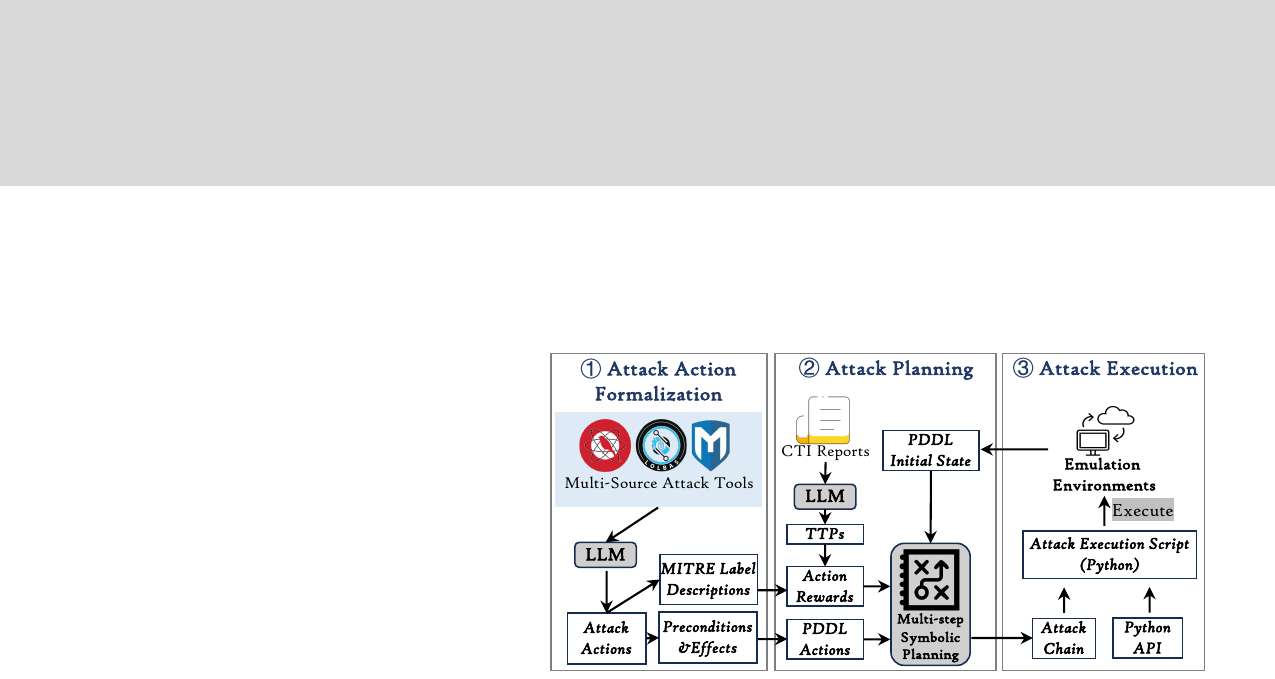}
    \caption{System overview.}
    \label{fig:overallframework}
    \vspace{-0.35in}
\end{figure}

Based on the attack actions and AALM, we design \SysName, an automated, customizable, causality-preserving cyberattack emulation system.
% As shown in Figure~\ref{fig:overallframework}, \SysName extracts attack actions from the documentation of existing attack tools.
% With these actions, \SysName can automatically construct customized attack chains based on emulation environments and CTI reports.
% It eventually generates a Python script to execute the attacks in the emulation environments.
As shown in Figure~\ref{fig:overallframework}, \SysName\ encompasses three stages: \ding{172}\textit{Attack Action Formalization}, \ding{173}\textit{Attack Planning}, and \ding{174}\textit{Attack Execution}.
\SysName first builds attack actions from existing tools.
During attack planning, \SysName considers available attack actions, emulation environments, and CTI reports to formalize the planning problem in PDDL.
The symbolic planner then chains the actions to a plan.
Finally, a Python script is generated to execute the attack plan in the emulation environment.

\vspace{-0.2in}

\subsection{Attack Action Formalization}\label{sec:attack-tool-analyzer}
The emulated attacks leverage existing tools rather than zero-day exploits, ensuring that \SysName evaluates defense systems only against known attack techniques (we discussed ethical considerations in \S\ref{sec:ethical}).
% Additionally, by incorporating various tools used by real-world attackers, \SysName generates actionable attacks, increases the attack diversity and fidelity compared to the previous work~\cite{depasquale_chainreactor,applebaum2016intelligent,miller2018automated}.
\SysName reviews documentation of attack tools and converts them into structured attack actions defined in \S\ref{sec:linking}.
However, analyzing heterogeneous documentation is challenging.
Although some simple features, such as name and description, are provided in the documentation and can be extracted with regular expressions, identifying preconditions and effects requires a deeper semantic understanding and substantial domain expertise in cybersecurity.

\begin{figure*}
    \centering
    \includegraphics[width=0.92\linewidth]{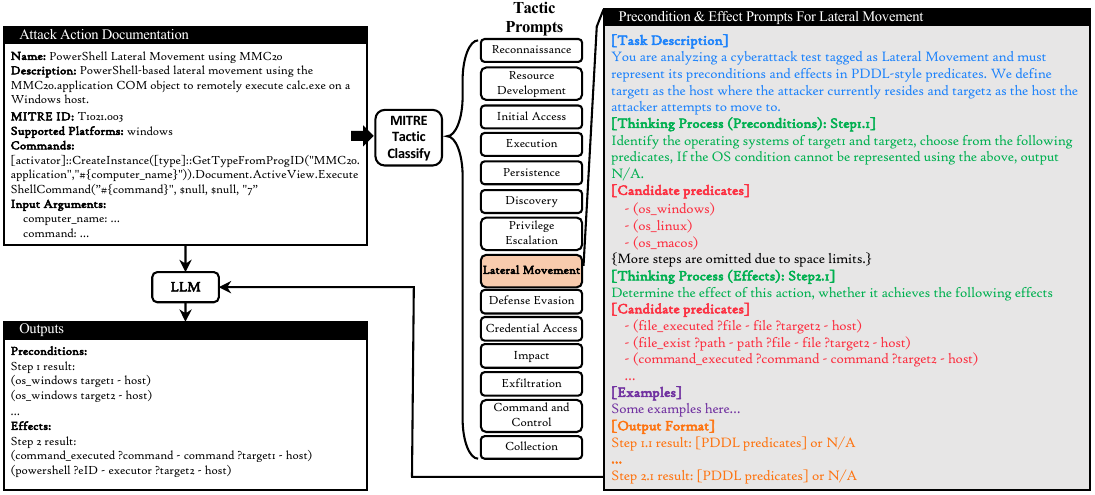}
    \caption{An example showing how \SysName analyzes preconditions and effects of an attack action using AALM.}
    \label{fig:prompts_engineering}
    \vspace{-0.1in}
\end{figure*}

Recent studies show that LLMs demonstrate strong capabilities in both text comprehension and domain-specific reasoning~\cite{wei2022chain,hendrycks2020measuring}.
\SysName leverages LLMs to formalize third-party tools into attack actions.
As shown in Figure~\ref{fig:prompts_engineering}, we begin by categorizing all tools into 14 groups based on their corresponding MITRE tactics.
For each tactic, we design a prompt for LLM to analyze the preconditions and effects.
The prompt consists of five components: a task description, a step-by-step thinking process (a series of questions), a set of candidate predicates from the AALM, illustrative examples of correct outputs and common mistakes, and formatting requirements for the output.
For each question in the thinking process, the LLM is restricted to selecting only from the given predicate set.
If no suitable predicate is available in AALM, the LLM is instructed to output \textit{N/A}, and the action is excluded from the attack emulation.
This approach minimizes hallucination and ensures that only actions compliant with AALM specifications are retained for subsequent symbolic planning.

As an example, consider tools used for payload generation, which fall under the \textit{Resource Development} tactic, including widely used tools such as Msfvenom, Sliver, and Empire.
The corresponding prompt first outlines the function of these tools and the analysis task.
It then details the analysis process.
For preconditions, the LLM is asked to identify the target OS and architecture of the payload and selects matching predicates from AALM; if none apply, it returns N/A.
For effects, the thinking process begins by checking whether the payload type matches any predefined payload types in AALM; if not, N/A is returned.
Next, it determines the payload format (file, shellcode, one-liner command, raw packet).
If the payload is a file, its file type is also identified.
If any of these steps yield N/A, \SysName excludes the action from the action space.
In our experiments, \SysName uses GPT-o3 (o3-2025-04-16) as the underlying LLM for predicate extraction.
Prompts were iteratively refined during development using a small validation set for each tool category to ensure syntactic correctness.
Once finalized, the prompts were fixed and not modified during evaluation.

Experimental results show that AALM and our prompt engineering can cover the majority of attack actions and formalize more than 5000 actions.
AALM also addresses the problem of relying on MITRE tactic labels in attack emulation.
For instance, certain tests in Atomic Red Team are labeled as \textit{Lateral Movement}, yet their actual effect does not achieve remote file transfer or code execution.
Therefore, \SysName ensures these actions are not included in the emulation solely based on their MITRE labels.
In \S\ref{sec:exp-correctness}, we show how this helps avoid invalid attack emulation compared to existing work.

% We reduce hallucination by limiting the LLMs to select only from a fixed set of predefined predicates included in the prompt.
% Enhanced by the prompt engineering and AALM, the experiment results in \S\ref{sec:exp-correctness} show that the overall attack success rates surpass those of leading attack emulation work.
% Although the prompt engineering requires some initial manual effort, this is a one-time step completed during the development of \SysName.
% As a result, the overall process is fully automated from the user’s perspective and requires no manual intervention at runtime.
% This paper focuses on defining AALM and automating action formalization using off-the-shelf LLMs; advancing LLM capabilities is beyond our scope (see \S\ref{sec:limitation-llm-acc} for more details).

\subsection{Attack Planning}\label{sec:design-attack-planning}
\SysName uses symbolic planning to chain the attack actions.
It represents attack actions and emulation environments in PDDL, which is then consumed by an off-the-shelf symbolic planner to generate execution plans.
Moreover, we designed a reward function mechanism to customize attack emulation based on CTI reports and a multi-step planning process to generate multi-stage attack emulation plans.
\subsubsection{Representing Emulation Environments in PDDL}\label{sec:starting-state}
The emulation environment, i.e., the target machine(s) of the attack, determines which actions are feasible.
Therefore, the starting state $I$ in attack planning should reflect the initial state of the emulation environment.
Prior to emulation, basic information about the target machine(s) is required, including operating systems, known vulnerabilities, installed software, and domain or network configurations.
This information can be easily obtained by running automated asset-inventory tools such as OpenSCAP, OpenVAS, and osquery~\cite{openscap,openvas,osquery}.
\SysName supports parsing the outputs of these tools and automatically map them to the environment predicates in AALM.
In \S\ref{sec:customization-env}, we will show how \SysName generates customized attacks based on different emulation environments.

\subsubsection{Reward Function for CTI-based Customization}\label{sec:attack-report-analyzer}
\SysName aims to emulate customized cyberattacks tailored to the characteristics of specific threat groups.
To achieve this, \SysName leverages CTI reports, which document the behaviors of real attackers.
We design the reward function to generate attack chains that closely resemble CTI reports.
We first extract TTPs from the reports.
Despite various approaches in analyzing CTI reports~\cite{ttpdrill,extractor,li2022attackg,alam2023looking,rahman2023attackers,kumarasinghe2024semantic}, our experiment in~\ref{app:additional_exp} shows that LLMs surpass these methods in terms of the accuracy.
Therefore, we employ off-the-shelf LLMs to extract TTPs from the reports.
To mitigate the hallucination issue and standardize the output, we include MITRE ATT\&CK TTP definitions and an output template in the prompt.
The improvement of CTI report analysis is not the focus of this paper.

\SysName customizes attack emulation based on CTI reports by generating attack chains that best match the TTPs extracted from the report.
To accomplish this, we assign a reward to each attack action.
Attack actions whose TTPs appear in the report are assigned higher rewards, while others receive lower rewards.
By doing so, maximizing the total reward of an attack chain directly corresponds to maximizing its similarity to the CTI report.
% Moreover, some symbolic planners drives the planner to favor the action chain that maximizes cumulative rewards.
To compute the reward of an action, we first check whether its MITRE label appears in the TTP set extracted from the report.
If not, the reward is set to zero.
Otherwise, the reward is computed as the cosine similarity between the embeddings of the action description and that of the corresponding TTP description.
Therefore, actions with more similar semantics will receive higher rewards.
For example, consider two privilege escalation actions: bypassing User Account Control (UAC) and process injection.
While both actions can elevate the attacker’s privileges, if the report contains some phrases such as \textit{abusing the UAC}, the former action will have a higher reward.
If two attack chains are generated in the same emulation environment, the one with the higher reward is selected.
Please note that we deliberately chose cosine similarity for its simplicity, efficiency, and ease of integration into planning.
More sophisticated semantic alignment functions (e.g., cross-encoders or entailment-based models) can be easily plugged in.
The evaluation in \S\ref{sec:customization-cti} shows that the cosine-based rewards yield over 30\% improvement in CTI-level customization.
% We leave the implementations and evaluations of more alternatives to future work.

\subsubsection{Multi-step Symbolic Planning}\label{sec:goal-state}
Setting the goal states of the planning problem remains challenging for the multi-step attacks.
On one hand, setting a single predicate as the goal state results in short chains that achieve only one attack objective.
However, if we combine all desired predicates into a single goal state, any inaccessible predicate causes the entire planning process to fail.
To address this, we designed a multi-step planning procedure.
Specifically, we created 12 pseudo actions, which correspond to 12 high-level attack stages defined by MITRE tactics.
The preconditions of a pseudo action are predicates that indicate the success of a given stage, connected by a logical \textit{or}.
The effect of each pseudo action is a dummy predicate that serves as the planning goal for that stage.
% For example, the pseudo action for persistence represents achieving persistence on a target machine.
% Its preconditions include predicates related to scheduled file or command execution, valid access credentials, and other persistence-relevant conditions.
% Once any of these preconditions is satisfied, the pseudo action is considered complete, and persistence is achieved.
% The goal state in planning is set to the dummy effect predicate of the pseudo action.
The symbolic planner then searches for a feasible attack chain that can achieve this goal.
When multiple chains exist, \SysName selects the one with the highest reward.
Pseudo action provides a concise and clear way to organize attack stages in symbolic planning.
The multi-step planning is then conducted according to the order of the APT lifecycle: initial access, execution, discovery, persistence, privilege escalation, defense evasion, credential access, lateral movement, collection, command and control, exfiltration, and impact.
The goal state of each step is the dummy effect predicate of the pseudo action.
The initial state before the first attack stage consists of the emulation environment predicates.
After completing a stage, \SysName collects the effect predicates of the newly selected actions and adds them to the initial state.
With the updated initial state, \SysName proceeds to plan the attack path for the next stage.
For instance, after completing the execution stage, we obtain \textsf{(sliver\_session ?e-executor ?t-host)} as one of the effect predicates.
This predicate is then added to the initial state for planning the next stage.
Note that the attack stage order is customizable.
For instance, sabotage-focused attacks may skip data collection and exfiltration stages.
Users can also reorder certain stages, as the predefined planning sequence does not constrain the causal dependencies between actions.
For example, if persistence requires privilege escalation as a precondition, \SysName will automatically place privilege escalation actions earlier, even if their stage comes later in the original planning sequence.
Multi-step planning of \SysName maximizes attack tactic coverage and stays robust when some stages are infeasible.

\textbf{Handling Execution Failures:} 
\SysName explicitly handles execution-time failures during attack emulation. When an action fails to execute due to unexpected runtime errors, \SysName removes the failed action from the action space and triggers re-planning for the current attack stage.
Since attack campaigns are decomposed into 12 stages, execution failures are isolated within the current stage and do not affect previously completed stages.
In other words, re-planning is restricted to the stage at which the failure occurs.
For example, if a PowerShell-based data collection action fails during execution, \SysName excludes this action from the action space and re-plans the data collection stage using the remaining candidate actions.
If no alternative action sequence can achieve the objective of the current stage, \SysName skips the stage and proceeds to plan subsequent stages.
This design ensures that execution failures do not halt the entire attack emulation process and allows \SysName to adaptively recover from runtime errors without requiring global rollback.

\subsection{Attack Execution}
The final step is to execute the generated attacks in the emulation environment.
Different from existing work~\cite{depasquale_chainreactor,applebaum2016intelligent,miller2018automated,holm2022lore}, \SysName can generate actionable attack emulation plans since each attack action corresponds to a concrete instruction from a tool.
To further enhance automation, we developed a Python function API covering all tools integrated in \SysName, including running Metasploit modules, executing commands in the Sliver console, executing commands on the attacker machine, and simulating simple user actions.
Based on the action sequence produced by the planner, \SysName automatically generates a Python script using this API.
This allows an entire attack chain to be executed by simply running a script.
% Another benefit of actionable attacks is that users can choose different data collectors to gather attack traces.
In \S\ref{sec:E6-real-defense}, we collected system logs when executing emulated attacks and used them to test three advanced intrusion detection systems.

%% file: sections/Evaluation.tex
We evaluate the performance of \SysName along five key dimensions.
Specifically, we assess whether \SysName can (1) successfully emulate attacks,
(2) generate causally coherent attack chains,
(3) customize attack emulation based on target environments and CTI reports,
(4) improve the scale and diversity of attack emulation, and
(5) support the evaluation and benchmarking of security defenses.
We present a case study to further illustrate how \SysName helps generate a complete, causality-preserved attack chain.
More evaluations, including the costs of \SysName and the accuracy of TTP extraction, are presented in Appendix~\ref{app:additional_exp}.

\vspace{-0.1in}
\subsection{Evaluation Setup}\label{sec:dataset}

\textbf{Attack Tools.} 
\SysName converts existing attack tools to attack actions.
In this paper, we utilize Atomic Red Team \cite{noauthor_atomicredteam_nodate}, Metasploit~\cite{metasploit}, Meterpreter~\cite{meterpreter}, Sliver~\cite{sliver}, and MSFvenom~\cite{msfvenom}.
They represent typical cyberattack tooling: Metasploit and MSFvenom cover common pre‑exploitation tasks, Atomic Red Team supplies the post‑exploitation library, and Meterpreter and Sliver are mature real‑world tools.
The above attack tools provide \SysName with 5,555 attack actions, more than many other automated cyberattack systems~\cite{deng2023pentestgpt,depasquale_chainreactor}.

\noindent\textbf{CTI Reports.} We collected 250 CTI reports from multiple sources~\cite{cisa_report,symantec_report,li2022attackg}.
The CTI report dataset used in our evaluation will be open-sourced.

% \begin{table}[h!]\footnotesize
% \centering
% \begin{tabular}{cc|c}
% \hline
% \multicolumn{2}{|c|}{Avg}                                                         & xxx \\ \hline
% \multicolumn{1}{|c|}{\multirow{3}{*}{Techniques in Frameworks}} & Atomic Red Team & x   \\ \cline{2-3} 
% \multicolumn{1}{|c|}{}                                          & MetaSploit      & x   \\ \cline{2-3} 
% \multicolumn{1}{|c|}{}                                          & LOLBAS          & x   \\ \hline
% \end{tabular}
% \end{table}

% \begin{table}[htbp]
%     \footnotesize
%     \centering
%     \vspace{-0.05in}   
%     \caption{}
%     \vspace{-0.1in}
%     \begin{tabular}{c|c|c}
%     \toprule
%      \multicolumn{2}{c|}{Poirot} &  \\ \midrule
%     Precision &Recall &Precisio \\  \midrule
%     Precision &Recall &Precisio \\
%     \bottomrule
%     \end{tabular}
%     \vspace{-0.05in}   
%     \label{tab:comparsion}
% \end{table}

% \begin{table}[h!]\footnotesize
%     \centering
%     \caption{dataset}
%     \label{tab:dataset}
%     \begin{tabular}{cc|c}
% \toprule
%  & Vanilla GPT-4o & Attack Range \                      \\ \midrule
%  Correct Syntax & 21/22 &     22/22  \\
%  Basic Functionality & 0/22 &     22/22   \\
%  Basic Functionality & 0/22 &     22/22 \\
%  \bottomrule
% \end{tabular}
% \end{table}

\noindent\textbf{Implementations.}
We implemented \SysName\ with 4K LoC in Python.
We used the GPT models from OpenAI and leveraged Fast Downward~\cite{fast_downward} to solve symbolic planning problems.
We set up the attack emulation environment infrastructure using Oracle VirtualBox, creating a small-scale LAN with 15 different hosts.
We release the code, datasets, and results to the public to facilitate further research.

\noindent\textbf{Baselines.}
As shown in Table~\ref{tab:baseline_comparison}, the baselines can be divided into four categories.
The first category is human-crafted cyberattack emulation datasets, including MITRE Evaluation~\cite{citd_eval}, and APT Attack Simulation~\cite{apt_attack_simulation}, a popular open-source attack sets.
The second category consists of procedural attack orchestration frameworks, including Caldera~\cite{caldera}, Attack Range~\cite{attackrange}, and PurpleSharp~\cite{purplesharp}.
The third category is the single-point testing tools, including Atomic Red Team~\cite{atomicredteam}, Metasploit~\cite{metasploit}, and Viper~\cite{viper}.
The last category is the abstract attack modeling studies~\cite{holm2022lore,applebaum2016intelligent,miller2018automated,depasquale_chainreactor}.
Specifically, when evaluating the success rate and causality preservation of attacks, we compared \SysName against PurpleSharp, Caldera, and the MITRE Evaluation because they provide modularized attacks, which are well-structured and suitable for quantitative analysis.

\vspace{-0.1in}
\subsection{Success Rate of Emulated Attacks}\label{sec:exp-correctness}
We first evaluated whether the attacks could successfully achieve malicious objectives.
We examined the main objectives of cyberattacks~\cite{apt-101-1,apt-101-2}, including espionage, data theft, sabotage, and gaining a long-term strategic advantage, and mapped them to six MITRE tactics: persistence, privilege escalation, collection, exfiltration, impact, and lateral movement.
Note that the remaining tactics, such as initial access and execution, primarily correspond to intermediate stages of an attack.
We then evaluated how many emulated attacks successfully accomplished these tactics.
Specifically, for privilege escalation, we assessed whether the emulated attack gained elevated privileges (e.g., root or sudo on Unix systems, or Administrator and System on Windows).
For persistence, we evaluated whether the attack established persistent access by maintaining a C2 agent or obtaining valid credentials.
For collection, we verified whether the attack acquired user data.
For exfiltration, we checked whether the attack transmitted collected data back to the attacker.
For lateral movement, we measured whether the attack successfully moved laterally to another machine by executing a file, command, or script on a second host.
For impact, we assessed whether the attack caused any destructive effects, including deleting user files, removing access, stopping services, or shutting down the system.
The results in Table~\ref{tab:success_rate_comparison} show that \SysName generates almost 10 times more attacks than the baseline and achieves the highest success rate across all tactics.
We analyzed the attack chains and identified two main reasons for the superiority of \SysName.
First, \SysName uses AALM to link attack actions, rather than relying on MITRE labels to arrange actions.
This avoids the invalid step due to the mismatch between MITRE labels and the actual action effects.
For example, in MITRE's Fin7 emulation~\cite{mitre_fin7}, \textit{Dump SAM via Mimikatz(T1003.002)} was used for privilege escalation.
However, executing this step does not directly provide the attacker with administrative privileges.
It merely dumps SAM credential data; attackers must take additional steps to extract plaintext credentials or hashes from it and then leverage those to gain high-privilege access.
On the contrary, \SysName ensures the high-privilege access after the privilege escalation step because AALM provides more concrete and accurate definitions of effects, closely aligned with actual attack emulation, which enhances the success rate and realism of emulations.
Second, multi-step planning allows \SysName to integrate multiple malicious behaviors into a single attack chain.
In contrast, most attack chains from tools like PurpleSharp and Caldera focus on just one tactic.
With multi-stage planning, users can freely determine which tactics appear in the chain, enabling broader coverage of malicious behaviors within a single emulation scenario.

\vspace{-0.2in}
\begin{table}[h]
    \centering
    \scriptsize
    \caption{Comparison of success rates(with 95\% confidence intervals).}
    \label{tab:success_rate_comparison}
\begin{tabular}{l|r|r|r|r}
\toprule
                     & PurpleSharp & Caldera & MITRE Eval. & \SysName   \\
\midrule
\# Total Chains      & 23          & 28      & 8           & \textbf{250}      \\
\midrule
Persistence          & 8.7\% (1.5,27.0)      & 0 (0.0,12.3)      & 87.5\% (47.4,99.7)    & \textbf{95.6\%} (92.3,97.8)  \\
Privilege Esc. & 4.4\% (0.1,22.0)     & 3.6\% (0.1,18.4) & 50.0\%  (15.7,84.3)   & \textbf{100.0\%} (98.5,100.0) \\
Collection           & 0  (0.0,14.8)         & 39.3\% (21.5,59.4) & 62.5\% (24.5,91.5)    & \textbf{98.0\%} (95.4,99.4)  \\
Exfiltration         & 0 (0.0,14.8)          & 28.6\% (13.2,48.7) & 37.5\%  (8.5,75.5)   & \textbf{100.0\%} (98.5,100.0)        \\
Impact               & 0 (0.0,14.8)          & 7.1\% (0.9,23.5)  & 37.5\% (8.5,75.5)    & \textbf{59.2\%} (52.8,65.4)  \\
Lateral Move.    & 30.4\% (13.2,52.9)    & 28.6\% (13.2,48.7)  & 87.5\% (47.4,99.7)     &  \textbf{100.0\%} (98.5,100.0)
\\ \bottomrule
\end{tabular}
    \vspace{-0.2in}
\end{table}

% PDDL limitations and incomplete information can cause inaccuracies in generated chains.
% While AALM links many general actions, it lacks predicates for domain-specific cases like third-party services or cloud APIs, which require added knowledge.
% However, AALM is designed to be extensible.
% This limitation can thus be addressed by defining additional, domain-specific predicates.

\subsection{Causality in Emulated Attacks}\label{sec:exp-causality}
\SysName aims to emulate high-fidelity attacks that are not only successful, but also preserve causal relationships between actions.
For example, it is not sufficient to merely emulate successful data collection and exfiltration in isolation.
Additionally, what is exfiltrated should be exactly what was collected.
Similarly, the process that gains elevated privileges should be the one used in subsequent actions.
As illustrated in Figure\ref{fig:motivating_example}(a), although all four steps are completed successfully, they are not causally linked.
To quantify the causality in an attack chain, we assessed logical connections between steps from three dimensions:
1) \textit{Executor Establishment}: Whether preceding steps establish the executor to run the current step.
2) \textit{Argument Sharing}: Whether the step shares arguments (e.g., hostnames, file paths) with other steps in the chain.
3) \textit{Artifact Interaction}: Whether the step utilizes or affects system artifacts (e.g., files, processes) that are involved in other steps.
Correspondingly, we define three metrics: \textit{Executor Traceability Rate}, \textit{Shared Arguments Rate}, and \textit{Shared Artifacts Rate}.
Each metric measures the proportion of steps in a chain satisfying the corresponding criterion.
In addition, we evaluated the average length of the attack chains.

We verified attack chains from the baselines and \SysName, with the results summarized in Table~\ref{tab:attack_chain_comparison}.
Our analysis highlights three key findings.
First, \SysName and the MITRE Evaluations provide significantly longer attack chains than PurpleSharp and Caldera.
Longer chains can better reflect the nature of multi-step APT attacks and offer more opportunities for detection.
Please note that \SysName can generate even longer chains if increasing the number of actions per stage during the multi-step planning.
Second, \SysName and MITRE Evaluations can construct full-lifecycle attack chains.
We define a full-lifecycle attack as one commencing with initial access or execution and including at least one remaining downstream tactic.
In contrast, the chains from PurpleSharp and Caldera typically focus on a single tactic.
Finally, in terms of causality preserving, \SysName significantly outperforms PurpleSharp and Caldera, achieving comparable performance to the expert-crafted MITRE chains.
The executor traceability for both PurpleSharp and Caldera is zero.
This is an expected outcome of their common architecture: a sequence of isolated scripts orchestrated by a central control agent, without modeling how that agent is initially established on the victim machine.
In addition, the high rates of shared arguments and artifacts in \SysName's chains indicate strong internal connectivity.

\vspace{-0.15in}
\begin{table}[h]
    \centering
    \scriptsize
    \caption{Comparison of causality preservation.}
    \label{tab:attack_chain_comparison}
\begin{tabular}{c|c|c|ccc}
\toprule
\multicolumn{1}{l|}{\multirow{3}{*}{}} &
  \multicolumn{1}{c|}{
  \multirow{3}{*}{\makecell{Avg. \\ Steps}}} 
  % Avg.
  &
  \multirow{3}{*}{\makecell{Full \\ Lifecycle}} &
  \multicolumn{3}{c}{Causality-preservation} \\
\multicolumn{1}{l|}{} &  &
   &
  \multicolumn{1}{c}{\makecell{Executor \\ Trac.}} &
  \multicolumn{1}{c}{\makecell{Shared \\ Args}} &
  \multicolumn{1}{c}{\makecell{Shared \\ Artifact}} \\
\midrule
PurpleSharp & 3.2 & \xmark   & 0        & 0 & 20.3\% \\
Caldera     & 5.4 & \xmark   & 0        & 15.1\% & 54.6\% \\
MITRE Eval.  & 53.6 & \cmark & 94.4\% & 16.7\% & 93.4\% \\
\midrule
\SysName      & \textbf{55.2} & \cmark & \textbf{100\%}    & \textbf{40.0\%} & \textbf{99.4\%}
\\ \bottomrule
\end{tabular}
\vspace{-0.3in}
\end{table}

\begin{figure}
    \centering
    \includegraphics[width=0.5\linewidth]{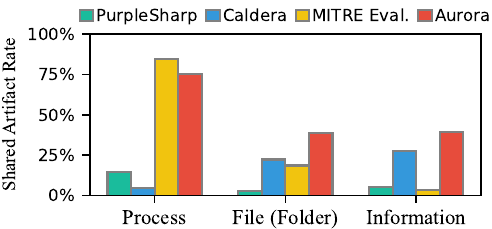}
    \caption{Comparison of shared artifact rates.}
    \label{fig:comparison-shared-artifact-rate}
    \vspace{-0.2in}
\end{figure}

To further analyze how AALM helps preserve causality, we examined the types of shared artifacts that connect steps in the emulated attack chains.
As shown in Figure~\ref{fig:comparison-shared-artifact-rate}, chains generated by \SysName exhibit a higher proportion of steps linked via shared files and information, while the MITRE chains show more connections through shared processes.
This is because \SysName models pre-foothold stages such as payload generation and initial access, where causal links are typically established via artifacts like files and target information.
In contrast, MITRE chains focus more on the stages after a foothold has already been established, where causality is more likely to manifest through process-level interactions.
Next, we investigated the contribution of different AALM predicate categories in preserving causality, since the linking relationship expressed by some categories in AALM (e.g., those for operating systems, CVEs, or exploit-payload mappings) has been considered in prior attack planning work~\cite{phillips1998graph,hu2020automated,chen2023gail,ghanem2023hierarchical}.
Our findings show that the novel predicate categories in AALM, such as those related to executors, payloads, files, information, users, and data, are crucial: they constitute 70.1\% of all predicate connections in the generated attack chains.

\subsection{Customization of Attack Emulation}\label{sec:ablation_cost_function}
% In this section, we evaluate whether \SysName can customize attack emulation by mimicking real-world adversaries described in CTI reports.
\subsubsection{Customization based on CTI Reports}\label{sec:customization-cti}
% We designed the reward function to prioritize the attack actions highlighted in the reports during planning.
% For this evaluation, we selected 50 CTI reports and generated two attack chains for each: a report-guided chain, using our reward-based planning tailored to the report, and a report‑agnostic chain created from the same candidate actions without CTI guidance.
% We then compare the number of CTI-mentioned actions present in the report-guided versus report‑agnostic chains.
% Figure~\ref{fig:E3-reward-effect} compares CTI-aligned actions in chains generated with (gold) and without (blue) reward guidance.
% The reward mechanism consistently guides \SysName to prioritize actions documented in the source CTI report, resulting in emulations that more accurately reflect reported adversary behavior.
% Sometimes the reward had no effect because either the reports' TTPs had no matching actions in our system, or the planner produced report‑aligned chains without reward by chance.

We designed the reward function to prioritize the attack actions mentioned in the CTI reports during planning to achieve CTI-level customization.
For this evaluation, we selected 50 CTI reports and generated two attack chains for each: a report-guided chain, using the reward functions during planning, and a report‑agnostic chain created from the same action space without CTI guidance.
We then compare the number of CTI-mentioned techniques in the generated attacks.
Across 50 CTI reports, incorporating the reward function during planning increases the number of CTI-mentioned techniques from 13.02 to 16.98 on average (+30.5\%). A paired t-test shows the improvement is statistically significant.

\begin{figure}[t]
    \centering
    \includegraphics[width=0.5\linewidth]{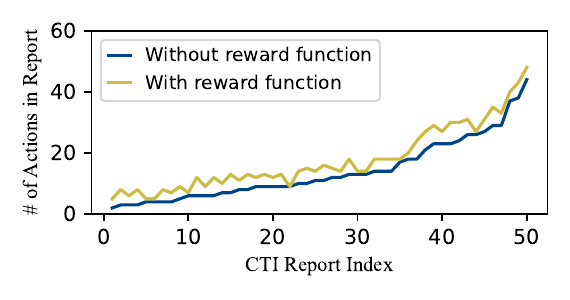}
    \caption{Fidelity of emulated attack chains to source CTI reports.}
    \label{fig:E3-reward-effect}
    \vspace{-0.2in}
\end{figure}

% Figure~\ref{fig:cost_function_impact} illustrates the impact of the reward function using two generated attack chains.
% Due to space constraints, other action details are omitted.
% Both chains achieve the same objective: obtaining a Meterpreter session on the target host and performing post-exploitation commands based on that.
% One way is exploiting the Apache Struts vulnerability (CVE-2016-3087) on the target host and executing a Meterpreter payload, which is chosen by the planner without the reward function.
% However, given the CTI report profiling a real attacker, Emotet Campaign, we observe that they typically gain initial access and execution through phishing email attachments and user interaction.
% As a result, the action \textit{User download and execute the phishing attachment} receives a higher reward and is thus prioritized during planning.
% Thus, the planner opts for this longer attack chain, which aligns better with the profile with a higher total reward.
% This example shows how \SysName emulates the specific attackers in CTI reports with the help of the reward function.

% \begin{figure}[tbp]
%     \centering
%     \includegraphics[width=\linewidth]{figures/cost_function_exp.pdf}
%     \caption{Comparison of two planned attacks with/without the cost function in the PDDL domain.}
%     \label{fig:cost_function_impact}
% \end{figure}

We also present a case study, a CISA report on Phobos Ransomware, to show how the reward function mechanism guides the planner toward report-aligned attacks to achieve report-level customization.
In the initial access stage, the report notes that \textit{``...threat actors send spoofed email attachments that are embedded with hidden payloads such as SmokeLoader...''}.
Therefore, \SysName selects the action of delivering payloads via email attachments (T1566.001) as malicious files (T1204.002) over other alternatives like using Metasploit exploit modules or supply chain attacks.
For the persistence, the report states that \textit{``...Phobos has also been observed using Windows Startup folders and Run Registry Keys to maintain persistence...''}.
Accordingly, \SysName favors the action that modifies the user shell folder startup value (T1547.001) over other options, such as creating a new account (T1136) or abusing boot or login initialization scripts (T1037).
% Notably, when faced with two available implementations for this action, one using cmd.exe and another using PowerShell, \SysName chooses the cmd.exe version.
% This choice is subtly influenced by another detail in the report stating that Phobos actors use cmd.exe to deploy payloads, demonstrating the reward function's ability to capture nuanced adversary preferences.
% Moreover, the reward function helps in navigating the extensive space of discovery and impact actions.
% For example, the report mentions that \textit{``Phobos ransomware actors...modifying system firewall configurations."} (T1562.004)
% Based on available actions and rewards, \SysName selects a closely related action: tampering with Windows Defender Registry settings (T1562.002).
Please note that \SysName balances CTI fidelity with actionability and causality.
When a tool mentioned in the report, like SmokeLoader, is unavailable, it substitutes a close alternative (e.g., Sliver) and arranges the subsequent actions automatically to preserve the attack chain’s logic.
% In summary, the goal of \SysName is not to perfectly replicate CTI reports verbatim, but to integrate intelligence from them to generate attack chains that are actionable, causality-preserved, and realistic within the context of the given environment and tools.

\vspace{-0.2in}
\subsubsection{Customization based on Emulation Environments}\label{sec:customization-env}
\begin{figure}
    \centering
    \includegraphics[width=\linewidth]{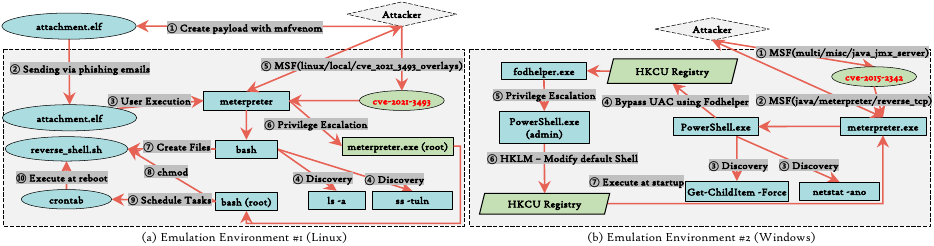}
    \caption{Customized attacks for different environments.}
    \label{fig:customization-env}
    \vspace{-0.2in}
\end{figure}

We demonstrate \SysName's capability of tailoring attacks to specific user environments through a case study where \SysName generates attack plans for two distinct environments: a modern Linux system (CVE-2021-3493) and an older Windows system (CVE-2015-2342).
For both environments, the objective was to create a five-stage attack plan covering initial access, execution, discovery, privilege escalation, and persistence. By simply modifying the initial state predicates in the respective PDDL problem files, \SysName's symbolic planner automatically selected compatible attack actions from our linking model to generate a unique chain for each environment. Figure~\ref{fig:customization-env} illustrates these results, showing two attack chains that achieve the same high-level goals using entirely different, environment-specific implementations.
Importantly, \SysName does not fail to generate attack chains due to the absence of environment-specific actions. When certain specialized actions are unavailable, \SysName can always fall back to composing attack chains using more generic and widely applicable actions, ensuring that valid multi-stage attack plans can still be generated for a given environment.

\vspace{-0.1in}

\subsection{Diversity of Emulated Cyberattacks}\label{sec:eval_ttp_coverage}
This section evaluates the scale and diversity of the attack chains generated by \SysName. We assess scale by the total number of generated attack chains and diversity by the number of unique attack actions, MITRE ATT\&CK tactics, and techniques covered. Our comparison includes a wide range of baselines.
As detailed in Table~\ref{tab:diversity}, \SysName surpasses all baselines in both scale and diversity.
In terms of scale, \SysName generates a significantly higher number of unique attack chains. This is a direct result of our approach, which defines and links modular attack actions, allowing for the creation of numerous valid chains through meaningful permutations.
In terms of diversity, \SysName covers the largest number of unique actions, techniques, and tactics. This breadth is achieved through its ability to integrate and leverage a wide array of third-party attack tools from different sources.

% Systems like ChainReactor lack specific attack commands and tools in their actions, relying instead on general conceptual actions, which further reduces their action number.

% The result shows \SysName\ enlarges the emulated attack technique space by 56.4\% and the emulated attack technique space by 41.4\%.
% The attacks from CTID are made up and released over five years, \SysName\ can generate these attack construction plans in just a few minutes with the help of LLM.
% Although the major goal of CTID is not to cover as many TTPs as possible, we believe the result shows the larger TTP coverage of \SysName to some extent.

\begin{table}[tbp]
    \centering
    \scriptsize
    \caption{Comparison of cyberattack dataset scale and TTP diversity.}
    \label{tab:diversity}
    \begin{tabular}{l|c|c|c|c}
    \toprule
& Attack	 & Attack	 & MITRE	 & MITRE \\
& Chains	 & Actions	 & Techniques	 & Tactics \\
\midrule
APT Simulation        & 13      & N/A$^1$           & N/A$^1$      & N/A$^1$ \\ 
MITRE Eval.        & 16      & 619           & 152       &12 \\  
PurpleSharp       & 25      & 181           & 30       &8  \\ 
Caldera       & 28      & 1530           & 309       &13  \\ 
Atomic Red Team       & N/A$^2$      & 1568           & 309       &13  \\ 
Viper        & N/A$^2$     & 97           & N/A$^3$       &10  \\
SVED        & N/A$^2$      & 4000+           & 262       &12  \\
Lore        & N/A$^3$      & N/A$^3$           & 68       &10  \\
ChainReactor       & N/A$^3$      & 31           & N/A$^3$       & N/A$^3$   \\
\midrule
\textbf{\SysName}        & 250      &      5555      & 327       & 14    \\
\bottomrule             
\end{tabular}
\\ \footnotesize \raggedright
$^1$ No modularized steps that can be mapped to MITRE ATT\&CK TTPs.\\
$^2$ Cannot construct multi-step cyberattacks.\\
$^3$ Not clearly reported.
    \vspace{-0.2in}
\end{table}

% \begin{figure}[tbp]
%     \centering
%     \includegraphics[width=0.98\linewidth]{figures/action_number_compare.pdf}
%     \caption{Comparison of the size of action space between \SysName and baselines.}
%     \label{fig:ttp_coverage}
% \end{figure}

\vspace{-0.2in}
\subsection{Evaluating Existing Detection Systems}\label{sec:E6-real-defense}
In this section, we used 250 attacks generated by \SysName to evaluate three advanced, open-source attack detectors, namely, \NODOZE~\cite{nodoze}, \PROVDETECTOR~\cite{provdetector}, and \FLASH~\cite{rehman2024flash}.
During these emulations, system traces were collected via ETW and converted into provenance graphs.
We established ground truth by labeling system entities involved in the attacks.
For training data, we simulated five hours of benign user activity. All three detectors perform anomaly detection on provenance graphs, but use different techniques.
We evaluated the true positive rates (TPR) and false positive rates (FPR).
For each metric, we first apply the Friedman test~\cite{friedman1937use} to assess overall differences among detectors.
When significant, we conduct post-hoc pairwise comparisons using the Wilcoxon signed-rank test with Holm correction $(\alpha = 0.05)$~\cite{wilcoxon1992individual,holm1979simple}.

\begin{table}[tbp]
\scriptsize
\centering
\scriptsize
    \caption{Comparison of detection performance against attacks generated by \SysName (with 95\% bootstrap confidence intervals).}
    \label{tab:detection-rates}
    \small
    \begin{tabular}{l|c|c|c}
    \toprule
        & \NODOZE               & \PROVDETECTOR         & \FLASH                \\ \midrule
    \multicolumn{1}{c|}{True Positive Rate} & 18.92\% (10.04,31.71)   & 48.16\%  (29.61,60.14)               & 9.67\%   (2.60,17.99)             \\
    False Positive Rate & 2.35\% (1.19,4.18)  & 1.72\% (0.92,2.58) & 0.06\% (0.01,0.12) \\
    \bottomrule
    \end{tabular}
    \vspace{-0.2in}
\end{table}

% Table~\ref{tab:detection-rates} shows the detection results.
% \NODOZE\ outperforms two other systems in the true positive rate while having more false alarms in these detection scenarios.
% Surprisingly, we observed significant drops in detection performance on our emulated attacks compared to the results on the benchmark datasets in the original papers, particularly in terms of true positive rates for \PROVDETECTOR\ and \FLASH.

Table~\ref{tab:detection-rates} shows the detection results.
The Friedman test indicates significant overall differences among detectors in both TPR ($p = 0.0046$) and FPR ($p = 0.0011$).
For TPR, our testing shows that \PROVDETECTOR\ and \NODOZE\ significantly outperform \FLASH.
\PROVDETECTOR\ also achieves higher TPR than \NODOZE, although the difference is not statistically significant.
For FPR, post-hoc analysis shows that \FLASH\ achieves a significantly lower false positive rate than both \PROVDETECTOR\ and \NODOZE, while the latter two are statistically indistinguishable.
In general, we observed a significant performance drop on our dataset compared to those used in prior work, likely due to the greater sophistication of attacks generated by \SysName.
We identified two primary reasons for this decline.
First, \FLASH’s GNN has only two hidden layers, limiting its receptive field to two-hop neighbors.
This may be sufficient for short attack paths, but inadequate for longer, causality-preserved attack chains from \SysName.
Capturing such paths would require a deeper GNN architecture.
Similarly, \PROVDETECTOR\ selects the \textit{K} rarest paths for anomaly detection.
However, since \SysName generates longer attack chains with more steps, the original \textit{K} setting is insufficient to cover all attack-related paths, leading to missed detections.
Second, \SysName heavily utilizes living-off-the-land (LotL) techniques, which blend in with benign behavior and may further reduce the effectiveness of anomaly-based detectors, particularly \NODOZE, which uses event frequency for detection.
Although a comprehensive benchmarking study lies outside the scope of this paper, this experiment demonstrates the necessity of building more frequently updated attack datasets.
% This evaluation demonstrates the key utility of \SysName: generating cyberattacks to benchmark defense systems and identify weaknesses.
% A comprehensive benchmarking study lies outside the scope of this paper.
% \SysName provides a valuable framework for security vendors to rigorously test and harden their defense systems against evolving, realistic threats.

\subsection{Case Study}
We present a case study on a recent real-world APT group active since 2024, Silver Fox~\cite{silver-fox-1,silver-fox-2}, to further illustrate how \SysName and AALM help generate a complete, causality-preserved attack chain.
At present, our understanding of the group remains limited.
Only a few isolated attack techniques are identified, which are insufficient for constructing attack emulations to support testing or data collection.
In collaboration with a red team, we obtained three steps used by the Silver Fox attack, and manually assigned preconditions and effects according to AALM.
These three steps focus solely on the tactics of execution, persistence, and impact.
Without AALM, it would be hard and time-consuming to integrate them with existing actions to build a complete APT attack emulation.

\begin{figure}[htbp]
    \vspace{-0.15in}
    \centering
    \includegraphics[width=0.7\linewidth]{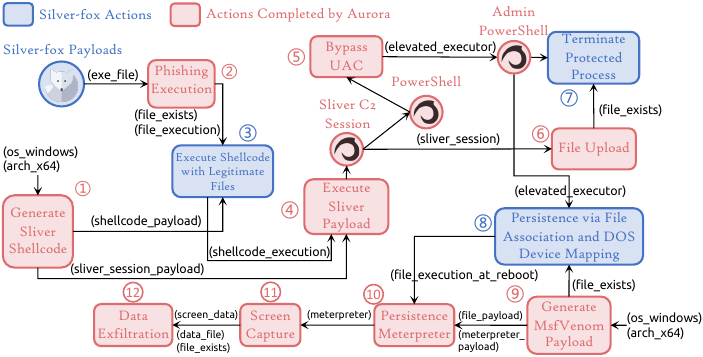}
    \caption{The case study showing how \SysName constructs a complete attack chain for three ``Silver Fox'' attack actions.}
    \label{fig:case-study}
    \vspace{-0.25in}
\end{figure}

The attack chain shown in Figure~\ref{fig:case-study} contains the three actions from Silver Fox and nine actions added by \SysName to form a complete, causality-preserved attack.
The first action shows Silver Fox is abusing legitimate, digitally signed programs to execute malicious code via techniques like DLL hijacking.
% The red team provides us with a legitimate executable to load and execute malicious shellcode.
% However, additional efforts are required to determine how to execute this executable and which shellcode would be needed to carry out the subsequent attack.
With \textsf{(file\_executed)} and \textsf{(shellcode\_payload)} as preconditions and \textsf{(shellcode\_executed)} as the effect, \SysName automatically built this action into the attack chain by linking it with phishing attachment execution and Sliver shellcode generation.
The second action exploits a known vulnerability in a legitimate, signed driver (wsftprm.sys) to gain kernel-level privileges.
% An attacker first loads the vulnerable driver and sends a specially crafted IOCTL command to the driver.
% This is often used to disable security software that is protected from user-mode termination.
This action requires the vulnerable driver file and the attacker executable to be placed \textsf{(file\_exists)}, and a PowerShell with administrative privileges \textsf{(powershell)} and \textsf{(elevated\_executor)}.
\SysName integrates it into a complete attack chain that uses Sliver to upload malicious files and perform a UAC bypass to obtain administrative privileges.
The third action of Silver Fox establishes stealthy persistence by chaining several evasive maneuvers.
% Upon reboot, the operating system places the file in the startup folder and automatically executes it, achieving persistence while simultaneously cleaning up the related files.
% Despite its complexity, AALM offers a straightforward way to link this action with others.
The precondition requires an administrative PowerShell environment to execute the script, represented by \textsf{(powershell)} and \textsf{(elevated\_executor)}.
The effect of the action is \textsf{(file\_execution\_at\_reboot)}.
The administrative PowerShell obtained for the previous action is reused. 
This action is then used to establish a persistent Meterpreter session for subsequent data exfiltration.
\vspace{-0.1in}

\subsection{Ablation Study}
We evaluate the effectiveness of our prompt design for predicate extraction by comparing LLM outputs against manually curated ground truth.
We perform stratified random sampling of 150 actions across 14 tactic groups.
Two authors independently annotate each action’s preconditions and effects based on AALM, resolving disagreements through discussion.
On this set, \SysName achieves high accuracy in predicate analysis.
Specifically, for 93\% of the sampled actions, the LLM correctly identifies preconditions that exactly match the ground truth, and for 86\% of the actions, the extracted effects match the ground truth.
% These results indicate that the proposed prompt design is effective in guiding the LLM to summarize the preconditions and effects.
We further analyze cases where the LLM outputs deviate from human annotations and identify the main sources of errors. First, some errors arise from application-specific dependencies between actions that are not explicitly defined in AALM.
For example, certain Meterpreter commands such as Drop Token require a prior Steal Token action, and KeyStrokeDump requires KeyStrokeCapture to be enabled beforehand.
Since these dependencies are tool-specific and not encoded in the current AALM predicate set, the LLM cannot reliably infer the correct preconditions and effects.
% Second, application-specific effects may be missing from AALM.
% For instance, actions that stop or start packet filtering mechanisms exhibit effects that are highly dependent on the packet filtering application.
Given the diversity of tools and applications involved in real-world attacks, it is impractical for AALM to predefine predicates for every specific application.
Nevertheless, AALM is designed to be extensible: users can introduce new predicates to capture such dependencies, enabling the LLM to recognize and formalize these actions accordingly.
Second, inaccuracies occur when a single complex action spans multiple tactics and produces multiple effects. In such cases, the LLM may fail to capture all effects accurately.
This limitation is not unique to AALM or LLM-based extraction; similar challenges arise when modeling attack processes using MITRE TTPs, where a single procedure may serve multiple tactical purposes.
% Overall, these error cases primarily reflect the current modeling scope of AALM rather than deficiencies in the LLM or prompt design.
Importantly, \SysName conservatively excludes actions whose predicates cannot be confidently formalized, preventing such inaccuracies from propagating into downstream symbolic planning.

We further evaluate the impact of the prompt engineering.
In \SysName, we design tactic-specific prompts that explicitly guide the LLM through a reasoning process for identifying preconditions and effects.
As an ablation baseline, we remove the prompt and instead provide the LLM with the complete set of predicates in AALM, allowing it to independently decide which predicates apply to a given action.
The results show that, compared to the prompt-engineered version, the ablated setting produces inaccurate preconditions and effects for 58\% of the evaluated actions, indicating that the LLM fails to exhaustively identify relevant conditions or effects when not explicitly guided by the prompt.
% These findings demonstrate that prompt engineering plays a critical role in ensuring accurate and complete predicate extraction.

%% file: sections/Discussion.tex
% \subsection{Scalability Issue of PDDL in Planning Cyberattacks}
% Although experimental results demonstrate that \SysName and AALM are effective in constructing attack emulation, we discussed their limitations in \S\ref{sec:exp-correctness}.
% The primary limitation lies in the scalability of PDDL-based symbolic planning, which restricts the number of predicates available to describe action preconditions and effects.
% However, the inherent complexity of cyberattacks results in a vast and diverse range of such preconditions and effects.
% To maintain a manageable predicate set, we must abstract or omit certain details, which can lead to inaccuracies in the generated attack plans.
% Extra work is needed to improve the scalability of symbolic planning, given the complexity of cyberattacks.

\vspace{-0.1in}
\textbf{More Advanced Attack Linking with LLMs.}\label{sec:limitation-llm-acc}
We leveraged LLMs to analyze the preconditions and effects of attack actions based on raw documentation and AALM.
We applied prompt engineering, injecting the step-by-step thinking process, illustrative examples, and candidate predicates into the prompt.
A key future research direction is enabling LLMs to link actions automatically.

\noindent\textbf{Apply AALM in Automated Penetration Testing.}
AALM links attack actions to enable symbolic planning for attack automation; the tools, actions, and predicates in \SysName are reusable for pentesting.
However, to realize automated penetration testing with AALM, several additional challenges must be addressed, such as incomplete environment knowledge, non-deterministic effects, unmodeled predicates, and the need to convert observed information into PDDL format.
\vspace{-0.1in}

%% file: sections/Conclusion.tex
\vspace{-0.1in}
In this paper, we define attack actions and the attack action linking model to integrate heterogeneous attack tools into causality-preserved cyberattacks.
We then introduce \SysName, the first automated, customizable, and causality-preserving cyberattack emulation system with LLM-powered symbolic planning.
The evaluation demonstrates that \SysName can emulate customized, causality-preserved attack chains with high scalability and diversity.
% We have published a cyberattack dataset containing over 250 attack chains.
\vspace{-0.1in}

%% file: sections/Ethical.tex
We take the ethical issues in attack emulation seriously.
The attack tool analyzer and attack report analyzer can only analyze existing attack tools and reports without involving any zero-day attacks developed.
Our goal is to leverage existing public knowledge of cyberattacks to enhance defensive capabilities.

%% file: sections/Appendix.tex
\section{Appendix}
\label{sec:appendix-pddl-problem-example}
\subsection{Searching for Multiple Attack Chains using Symbolic Planners}\label{appendix:pddl-searching}
To find as many attack chains as possible, we leverage the cost mechanism in symbolic planning.
Initially, all actions have a cost of zero, and the algorithm searches for the plan with the lowest total cost in this action space (since all plans have a total cost of 0 at this stage, the search returns the first attack plan it discovers).
After discovering a plan, we increase the costs of all actions within that plan by a fixed amount and update the domain.
Then we search for the lowest-cost plan in the updated domain. Because the actions from the first plan now have higher costs, the planner favors creating new plans using previously unused actions.
This iterative process continues until either no new plans are generated for several consecutive rounds or we reach our target number of generated plans.

\subsection{Supplementary Experimental Results}\label{app:additional_exp}
\textbf{Costs of \SysName:}
We assessed the time and monetary costs of \SysName.
The results are shown in Table~\ref{tab:time_cost}.
With LLMs, extracting attack actions from documentation and analyzing their preconditions and effects takes less than 10 seconds on average.
The time required to analyze one CTI report for attack customization is approximately 40 seconds.
Planning time varies with the length of the attack chain, taking approximately 0.5 to 2 hours for a 50-step chain.
Executing an emulated attack takes ten minutes on average.
We also evaluate the cost of using LLM.
The result shows that \SysName\ is economical in terms of both time and money.

\begin{table}[t]\footnotesize
    \centering
    \scriptsize
    \caption{Average Time and Monetary Costs of \SysName}
    \label{tab:time_cost}
    \begin{tabular}{c|c|c|c}
    \toprule
                       & \multirow{2}{*}{Time} & \multicolumn{2}{c}{LLM Costs}                                   \\
                       &         & Tokens      &   Cost (USD)                                    \\ \midrule
Attack Tool Analyzing $^1$        &  9.8s     & 10.8K          & 0.029                \\
% Attack Tool Analyzing - MITRE Info $^1$       & 8.2s      & 6.52K         & 0.018                \\
CTI Report Analyzing $^2$ &    26.1s   & 16.8K           & 0.055                \\
Reward Embedding $^2$     & 14.9s       & 95.945K           & 0.010 \\
Attack Planning $^3$       & 0.5-2h      & -     & -    \\
% Environment Setup $^3$       & 20.0s      & 0     & 0    \\
Attack Execution $^3$       & 9.3m      & -     & -    \\
\bottomrule             
\end{tabular}
\\ \footnotesize \raggedright
$^1$ per attack action.
$^2$ per report.
$^3$ per attack chain (50 actions on avg.).
\end{table}

\begin{table}[tbp]\footnotesize
    \centering
    \scriptsize
    \caption{The Performance Comparison of Attack Technique Extraction between \LADDER\ and \SysName}
    \label{tab:tt_extraction_acc}
    \begin{tabular}{c|rr|rr|rr}
    \toprule
\multirow{2}{*}{id} & \multicolumn{2}{c}{Recall} & \multicolumn{2}{c}{Precision} & \multicolumn{2}{c}{Correctness} \\
% \cmidrule{2-7}
 & Base          & Ours           & Base &  Ours          & Base & Ours           \\
\midrule
1                    & 61.5\%          & \textbf{100.0\%} & 42.1\% & \textbf{100.0\%} & 52.6\% & \textbf{84.6\%}  \\
2                    & 28.6\%          & \textbf{71.4\%}  & 17.4\% & \textbf{100.0\%} & 30.4\% & \textbf{100.0\%} \\
3                    & 20.0\%          & \textbf{100.0\%}  & 5.6\%  & \textbf{28.6\%}  & 22.2\% & \textbf{80.0\%}  \\
4                    & 28.9\% & \textbf{88.1}\%           & 43.3\% & \textbf{97.2\%}  & 43.3\% & \textbf{100.0\%} \\
5                    & 33.3\%          & \textbf{93.3\%}  & 41.7\% & \textbf{100.0\%} & 46.2\% & \textbf{100.0\%} \\
6                    & 58.3\%          & \textbf{62.5\%}  & 35.9\% & \textbf{93.8\%}  & 43.6\% & \textbf{100.0\%} \\
7                    & 17.1\%          & \textbf{38.3\%}  & 25.0\% & \textbf{78.6\%}  & 42.9\% & \textbf{80.0\%}  \\
8                    & 19.0\%          & \textbf{76.2\%}  & 25.0\% & \textbf{88.9\%}  & 37.5\% & \textbf{94.4\%}  \\
9                    & 25.0\%          & \textbf{100.0\%}  & 11.8\% & \textbf{53.8\%}  & 35.3\% & \textbf{84.6\%}  \\
10                   & 46.2\%          & \textbf{84.6\%}  & 22.2\% & \textbf{84.6\%}  & 33.3\% & \textbf{100.0\%} \\
\bottomrule
\end{tabular}
\end{table}

\noindent \textbf{Extracting TTPs from CTI Reports:} We evaluate the accuracy of extracting TTPs from CTI using reports from CISA~\cite{cisa_report} as the ground truth.
We employ \LADDER~\cite{alam2023looking} as the baseline.
% We calculated precision and recall for TTP extraction, alongside correctness, which measures the quality of the generated descriptions.
As shown in Table~\ref{tab:tt_extraction_acc}, the LLM-based report analyzer of \SysName\ achieves higher precision than the baseline, nearing or exceeding 80\% for almost all reports.
And the average recall is 69\%.
The result also shows that \SysName\ provides a more accurate description for each individual technique than \LADDER.
After examining the results, we identified two key advantages of using LLM.
First, LLMs have better capabilities in understanding long sentences.
Conversely, \LADDER\ tends to be distracted by irrelevant words in filenames or positions.
Second, LLM provides more comprehensible descriptions for the extracted TTPs.
In contrast, \LADDER\ sometimes yields incomplete sentences.

We also analyze the reasons for failure cases.
First, LLM makes mistakes when some descriptions are ambiguous.
For instance, a report mentions that \textit{affiliates communicate with victims via TOR, Tox, email, or encrypted applications}.
However, the LLM incorrectly identified this sentence as evidence of a command and control technique during the attack.
Second, LLMs sometimes fail to enumerate all TTPs exhaustively.
For instance, a report mentions that the attacker \textit{impersonates company IT and/or helpdesk staff to gain trust and obtain credentials}.
While the LLM successfully identifies Impersonation (T1656), it fails to recognize that this also demonstrates the attacker is gathering victim identity information (T1589).

%% file: refs.bib
@article{holm1979simple,
  title={A simple sequentially rejective multiple test procedure},
  author={Holm, Sture},
  journal={Scandinavian journal of statistics},
  pages={65--70},
  year={1979},
  publisher={JSTOR}
}

@incollection{wilcoxon1992individual,
  title={Individual comparisons by ranking methods},
  author={Wilcoxon, Frank},
  booktitle={Breakthroughs in statistics: Methodology and distribution},
  pages={196--202},
  year={1992},
  publisher={Springer}
}

@article{friedman1937use,
  title={The use of ranks to avoid the assumption of normality implicit in the analysis of variance},
  author={Friedman, Milton},
  journal={Journal of the american statistical association},
  volume={32},
  number={200},
  pages={675--701},
  year={1937},
  publisher={Taylor \& Francis}
}

@inproceedings{loevenich2025automating,
  title={Automating Cyber Threat Intelligence and Attack Chain Generation using Cyber Security Knowledge Graphs and Large Language Models},
  author={Loevenich, Johannes F and Adler, Erik and H{\"u}rten, Tobias and Spelter, Florian and Roncevic, Damian and Lopes, Roberto Rigolin F},
  booktitle={2025 International Conference on Military Communication and Information Systems (ICMCIS)},
  pages={1--10},
  year={2025},
  organization={IEEE}
}

@inproceedings{loevenich2025agentic,
  title={Agentic Generative AI for Automation of Cyber Security Attack Chains in Tactical MANETs},
  author={Loevenich, Johannes F and Lopes, Roberto Rigolin F},
  booktitle={2025 IEEE 50th Conference on Local Computer Networks (LCN)},
  pages={1--7},
  year={2025},
  organization={IEEE}
}

@article{hendrycks2020measuring,
  title={Measuring massive multitask language understanding},
  author={Hendrycks, Dan and Burns, Collin and Basart, Steven and Zou, Andy and Mazeika, Mantas and Song, Dawn and Steinhardt, Jacob},
  journal={arXiv preprint arXiv:2009.03300},
  year={2020}
}

@misc{cve,
    title = {CVE Program Mission},
    author = {The MITRE Corporation},
    key = {cve},
    howpublished = {\url{https://www.cve.org/}},
    urldate = {2025-08-01}
}

@misc{cpe,
    title = {Official Common Platform Enumeration (CPE) Dictionary},
    author = {National Vulnerabilities Database},
    howpublished = {\url{https://nvd.nist.gov/products/cpe}},
    urldate = {2025-08-01}
}

@inproceedings{sarraute2012pomdps,
  title={POMDPs make better hackers: Accounting for uncertainty in penetration testing},
  author={Sarraute, Carlos and Buffet, Olivier and Hoffmann, J{\"o}rg},
  booktitle={Proceedings of the AAAI Conference on Artificial Intelligence},
  volume={26},
  pages={1816--1824},
  year={2012}
}

@misc{openvas,
    title = {OpenVAS - Open Vulnerability Assessment Scanner},
    author = {OpenVAS},
    howpublished = {\url{https://www.openvas.org}},
    urldate = {2025-08-01}
}

@misc{openscap,
    title = {Open Source Security Compliance Solution},
    author = {OpenScap},
    howpublished = {\url{https://www.open-scap.org/}},
    urldate = {2025-08-01}
}

@misc{osquery,
    title = {osquery: a SQL powered operating system instrumentation, monitoring, and analytics framework.},
    author = {osquery},
    howpublished = {\url{https://github.com/osquery/osquery}},
    urldate = {2025-08-01}
}

@misc{apt-101-1,
    title = {What are advanced persistent threats?},
    author = {Gregg Lindemulder},
    howpublished = {\url{https://www.ibm.com/think/topics/advanced-persistent-threats?regionCode=US&languageCode=en&contactmodule=true&cm-history=US-en}},
    urldate = {2025-08-01}
}

@misc{apt-101-2,
    title = {What Is an Advanced Persistent Threat?},
    author = {Paloalto Network},
    howpublished = {\url{https://www.paloaltonetworks.com/cyberpedia/what-is-advanced-persistent-threat-apt?utm_source=chatgpt.com}},
    urldate = {2025-08-01}
}

@misc{silver-fox-1,
    title = {Silver Fox APT Targets Public Sector via Trojanized Medical Software},
    author = {Sila Özeren Hacıoğlu},
    howpublished = {\url{https://www.picussecurity.com/resource/blog/silver-fox-apt-targets-public-sector-via-trojanized-medical-software}},
    urldate = {2025-08-01}
}

@misc{silver-fox-2,
    title = {Silver Fox APT Attack Taiwan},
    author = {Charles Leigh},
    howpublished = {\url{https://westoahu.hawaii.edu/cyber/global-weekly-exec-summary/silver-fox-apt-attack-taiwan/}},
    urldate = {2025-08-01}
}

@misc{mitre_fin7,
    title = {Machine-Readable FIN7 Emulation Plan},
    author = {ATT&CK Evaluations},
    howpublished = {\url{https://github.com/attackevals/ael/tree/main/Enterprise/fin7/Emulation_Plan/yaml}},
    urldate = {2025-08-01}
}

@misc{msfvenom,
    title = {MSFvenom - Metasploit Unleashed},
    howpublished = {\url{https://www.offsec.com/metasploit-unleashed/msfvenom/}},
    key = {msfvenom},
    urldate = {2025-06-01}
}

@misc{cymulate,
    title = {Official website of Cymulate},
    key = {CYMULATE},
    howpublished = {\url{https://cymulate.com/}},
    urldate = {2025-06-01}
}

@misc{horizon3,
    title = {Are you vulnerable or exploitable? Official website of Horizon3},
    key = {horizon3},
    howpublished = {\url{https://horizon3.ai/}},
    urldate = {2025-06-01}
}

@misc{pentera,
    title = {Welcome to Pentera: Don’t assume. Validate. Because 'pretty certain' doesn't mean secure},
    key = {pentera},
    howpublished = {\url{https://pentera.io/}},
    urldate = {2025-06-01}
}

@misc{tibereu,
    title = {TIBER-EU FRAMEWORK: How to implement the European framework for Threat Intelligence-Based Ethical Red teaming},
    key = {TIBER-EU},
    howpublished = {\url{https://www.ecb.europa.eu/pub/pdf/other/ecb.tiber_eu_framework_2025~b32eff9a10.en.pdf}},
    urldate = {2025-05-01}
}

@misc{marketingreport,
    title = {Auromated Breach and Attack Simulation Market},
    key = {marketingreport},
    howpublished = {\url{https://www.marketsandmarkets.com/Market-Reports/automated-breach-attack-simulation-market-43164821.html}},
    urldate = {2025-05-01}
}

@inproceedings{alsaheel2021atlas,
  title={$\{$ATLAS$\}$: A sequence-based learning approach for attack investigation},
  author={Alsaheel, Abdulellah and Nan, Yuhong and Ma, Shiqing and Yu, Le and Walkup, Gregory and Celik, Z Berkay and Zhang, Xiangyu and Xu, Dongyan},
  booktitle={30th USENIX Security Symposium (USENIX Security 21)},
  pages={3005--3022},
  year={2021}
}

@misc{darpa_engagement,
    title = {DARPA Engagement Data},
    key = {darpa_engagement},
    howpublished = {\url{https://drive.google.com/drive/folders/1okt4AYElyBohW4XiOBqmsvjwXsnUjLVf}},
    urldate = {2025-04-01}
}

@misc{streamspot,
    title = {StreamSpot Data},
    key = {streamspot},
    howpublished = {\url{https://github.com/sbustreamspot/sbustreamspot-data}},
    urldate = {2025-04-01}
}

@article{ghanem2023hierarchical,
  title={Hierarchical reinforcement learning for efficient and effective automated penetration testing of large networks},
  author={Ghanem, Mohamed C and Chen, Thomas M and Nepomuceno, Erivelton G},
  journal={Journal of Intelligent Information Systems},
  volume={60},
  number={2},
  pages={281--303},
  year={2023},
  publisher={Springer}
}

@article{chen2023gail,
  title={GAIL-PT: An intelligent penetration testing framework with generative adversarial imitation learning},
  author={Chen, Jinyin and Hu, Shulong and Zheng, Haibin and Xing, Changyou and Zhang, Guomin},
  journal={Computers \& Security},
  volume={126},
  pages={103055},
  year={2023},
  publisher={Elsevier}
}

@article{li2024dynpen,
  title={DynPen: Automated Penetration Testing in Dynamic Network Scenarios Using Deep Reinforcement Learning},
  author={Li, Qianyu and Wang, Ruipeng and Li, Dong and Shi, Fan and Zhang, Min and Chattopadhyay, Anupam},
  journal={IEEE Transactions on Information Forensics and Security},
  year={2024},
  publisher={IEEE}
}

@inproceedings{holm2016sved,
  title={Sved: Scanning, vulnerabilities, exploits and detection},
  author={Holm, Hannes and Sommestad, Teodor},
  booktitle={MILCOM 2016-2016 IEEE Military Communications Conference},
  pages={976--981},
  year={2016},
  organization={IEEE}
}

@misc{cobaltstrike,
  author = {Fortra},
  title = {Software for Adversary Simulations and Red Team Operations},
  year = {2024},
  url = {https://www.cobaltstrike.com/},
note = {Accessed: 2025-04-01}
}

@inproceedings{jian2025,
	title={{ORTHRUS: Achieving High Quality of Attribution in Provenance-based Intrusion
	Detection Systems}},
	author={Jiang, Baoxiang and Bilot, Tristan  and El Madhoun, Nour and Al Agha, Khaldoun  and Zouaoui, Anis and Iqbal, Shahrear and Han, Xueyuan and Pasquier, Thomas},
	booktitle={Security Symposium (USENIX Sec'25)},
	year={2025},
	organization={USENIX}
}

@inproceedings{wang2020you,
  title={You Are What You Do: Hunting Stealthy Malware via Data Provenance Analysis.},
  author={Wang, Qi and Hassan, Wajih Ul and Li, Ding and Jee, Kangkook and Yu, Xiao and Zou, Kexuan and Rhee, Junghwan and Chen, Zhengzhang and Cheng, Wei and Gunter, Carl A and others},
  booktitle={NDSS},
  year={2020}
}

@article{han2020unicorn,
  title={Unicorn: Runtime provenance-based detector for advanced persistent threats},
  author={Han, Xueyuan and Pasquier, Thomas and Bates, Adam and Mickens, James and Seltzer, Margo},
  journal={arXiv preprint arXiv:2001.01525},
  year={2020}
}

@inproceedings{milajerdi2019poirot,
  title={Poirot: Aligning attack behavior with kernel audit records for cyber threat hunting},
  author={Milajerdi, Sadegh M and Eshete, Birhanu and Gjomemo, Rigel and Venkatakrishnan, VN},
  booktitle={Proceedings of the 2019 ACM SIGSAC conference on computer and communications security},
  pages={1795--1812},
  year={2019}
}

@inproceedings{zengy2022shadewatcher,
  title={Shadewatcher: Recommendation-guided cyber threat analysis using system audit records},
  author={Zengy, Jun and Wang, Xiang and Liu, Jiahao and Chen, Yinfang and Liang, Zhenkai and Chua, Tat-Seng and Chua, Zheng Leong},
  booktitle={2022 IEEE symposium on security and privacy (SP)},
  pages={489--506},
  year={2022},
  organization={IEEE}
}

@inproceedings{yang2023prographer,
  title={$\{$PROGRAPHER$\}$: An anomaly detection system based on provenance graph embedding},
  author={Yang, Fan and Xu, Jiacen and Xiong, Chunlin and Li, Zhou and Zhang, Kehuan},
  booktitle={32nd USENIX Security Symposium (USENIX Security 23)},
  pages={4355--4372},
  year={2023}
}

@article{li2023nodlink,
  title={Nodlink: An online system for fine-grained apt attack detection and investigation},
  author={Li, Shaofei and Dong, Feng and Xiao, Xusheng and Wang, Haoyu and Shao, Fei and Chen, Jiedong and Guo, Yao and Chen, Xiangqun and Li, Ding},
  journal={arXiv preprint arXiv:2311.02331},
  year={2023}
}

@inproceedings{cheng2024kairos,
  title={Kairos: Practical intrusion detection and investigation using whole-system provenance},
  author={Cheng, Zijun and Lv, Qiujian and Liang, Jinyuan and Wang, Yan and Sun, Degang and Pasquier, Thomas and Han, Xueyuan},
  booktitle={2024 IEEE Symposium on Security and Privacy (SP)},
  pages={3533--3551},
  year={2024},
  organization={IEEE}
}

@inproceedings{goyal2024r,
  title={R-caid: Embedding root cause analysis within provenance-based intrusion detection},
  author={Goyal, Akul and Wang, Gang and Bates, Adam},
  booktitle={2024 IEEE Symposium on Security and Privacy (SP)},
  pages={3515--3532},
  year={2024},
  organization={IEEE}
}

@inproceedings{jia2024magic,
  title={$\{$MAGIC$\}$: Detecting advanced persistent threats via masked graph representation learning},
  author={Jia, Zian and Xiong, Yun and Nan, Yuhong and Zhang, Yao and Zhao, Jinjing and Wen, Mi},
  booktitle={33rd USENIX Security Symposium (USENIX Security 24)},
  pages={5197--5214},
  year={2024}
}

@misc{meterpreter,
  author = {Rapid7},
  title = {Metasploit Documentation of Meterpreter},
  year = {2024},
  url = {https://docs.metasploit.com/docs/using-metasploit/advanced/meterpreter/meterpreter.html},
note = {Accessed: 2025-04-01}
}

@misc{sliver,
  author = {BishopFox},
  title = {Sliver},
  year = {2024},
  url = {https://github.com/BishopFox/sliver},
note = {Accessed: 2025-04-01}
}

@inproceedings{hoffmann2015simulated,
  title={Simulated penetration testing: From" dijkstra" to" turing test++"},
  author={Hoffmann, J{\"o}rg},
  booktitle={Proceedings of the international conference on automated planning and scheduling},
  volume={25},
  pages={364--372},
  year={2015}
}

@article{ding2023integrating,
  title={Integrating action knowledge and LLMs for task planning and situation handling in open worlds},
  author={Ding, Yan and Zhang, Xiaohan and Amiri, Saeid and Cao, Nieqing and Yang, Hao and Kaminski, Andy and Esselink, Chad and Zhang, Shiqi},
  journal={Autonomous Robots},
  volume={47},
  number={8},
  pages={981--997},
  year={2023},
  publisher={Springer}
}

@article{chen2024language,
  title={Language-augmented symbolic planner for open-world task planning},
  author={Chen, Guanqi and Yang, Lei and Jia, Ruixing and Hu, Zhe and Chen, Yizhou and Zhang, Wei and Wang, Wenping and Pan, Jia},
  journal={arXiv preprint arXiv:2407.09792},
  year={2024}
}

@misc{apt_attack_simulation,
  author = {S3N4T0R-0X0},
  title = {APT Attack Simulation},
  year = {2024},
  url = {https://github.com/S3N4T0R-0X0/APT-Attack-Simulation},
note = {Accessed: 2025-03-01}
}

@inproceedings{phillips1998graph,
  title={A graph-based system for network-vulnerability analysis},
  author={Phillips, Cynthia and Swiler, Laura Painton},
  booktitle={Proceedings of the 1998 workshop on New security paradigms},
  pages={71--79},
  year={1998}
}

@article{holm2022lore,
  title={Lore a red team emulation tool},
  author={Holm, Hannes},
  journal={IEEE Transactions on Dependable and Secure Computing},
  volume={20},
  number={2},
  pages={1596--1608},
  year={2022},
  publisher={IEEE}
}

@misc{fast_downward,
  author = {Artificial Intelligence Group - University of Basel},
  title = {Fast Downwards},
  year = {2024},
  url = {https://github.com/aibasel/downward},
note = {Accessed: 2024-11-01}
}

@article{wang2024incorporating,
  title={Incorporating Gradients to Rules: Towards Lightweight, Adaptive Provenance-based Intrusion Detection},
  author={Wang, Lingzhi and Shen, Xiangmin and Li, Weijian and Li, Zhenyuan and Sekar, R and Liu, Han and Chen, Yan},
  journal={arXiv preprint arXiv:2404.14720},
  year={2024}
}

@misc{viper,
  author = {FunnyWolf},
  title = {Viper},
  year = {2024},
  url = {https://github.com/FunnyWolf/Viper?tab=readme-ov-file},
note = {Accessed: 2024-11-01}
}

@inproceedings {depasquale_chainreactor,
author = {Giulio De Pasquale and Ilya Grishchenko and Riccardo Iesari and Gabriel Pizarro and Lorenzo Cavallaro and Christopher Kruegel and Giovanni Vigna},
title = {{ChainReactor}: Automated Privilege Escalation Chain Discovery via {AI} Planning},
booktitle = {33rd USENIX Security Symposium (USENIX Security 24)},
year = {2024},
isbn = {978-1-939133-44-1},
address = {Philadelphia, PA},
pages = {5913--5929},
url = {https://www.usenix.org/conference/usenixsecurity24/presentation/de-pasquale},
publisher = {USENIX Association},
month = aug
}

@inproceedings{kumarasinghe2024semantic,
  title={Semantic ranking for automated adversarial technique annotation in security text},
  author={Kumarasinghe, Udesh and Lekssays, Ahmed and Sencar, Husrev Taha and Boughorbel, Sabri and Elvitigala, Charitha and Nakov, Preslav},
  booktitle={Proceedings of the 19th ACM Asia Conference on Computer and Communications Security},
  pages={49--62},
  year={2024}
}

@article{rahman2023attackers,
  title={What are the attackers doing now? Automating cyberthreat intelligence extraction from text on pace with the changing threat landscape: A survey},
  author={Rahman, Md Rayhanur and Hezaveh, Rezvan Mahdavi and Williams, Laurie},
  journal={ACM Computing Surveys},
  volume={55},
  number={12},
  pages={1--36},
  year={2023},
  publisher={ACM New York, NY}
}

@inproceedings{applebaum2016intelligent,
  title={Intelligent, automated red team emulation},
  author={Applebaum, Andy and Miller, Doug and Strom, Blake and Korban, Chris and Wolf, Ross},
  booktitle={Proceedings of the 32nd annual conference on computer security applications},
  pages={363--373},
  year={2016}
}

@misc{symantec_report,
  author = {Symantec},
  title = {Symantec Enterprise Blogs Threat Intelligence},
  year = {2024},
  url = {https://symantec-enterprise-blogs.security.com/threat-intelligence/clasiopa-materials-research},
note = {Accessed: 2024-07-01}
}

@misc{cisa_report,
  author = {CISA},
  title = {Cybersecurity Alerts Advisories},
  year = {2024},
  url = {https://www.cisa.gov/news-events},
note = {Accessed: 2024-07-01}
}

@misc{citd_eval,
  author = {MITRE Engenuity},
  title = {CYBERSECURITY:
ATT\&CK® EVALUATIONS},
  year = {2024},
  url = {https://evals.mitre.org/},
note = {Accessed: 2025-06-01}
}

@misc{attackrange,
  author = {Splunk Threat Research Team},
  title = {Splunk Attack Range},
  year = {2024},
  url = {https://github.com/splunk/attack\_range},
note = {Accessed: 2024-07-01}
}

@inproceedings{li2022attackg,
  title={AttacKG: Constructing technique knowledge graph from cyber threat intelligence reports},
  author={Li, Zhenyuan and Zeng, Jun and Chen, Yan and Liang, Zhenkai},
  booktitle={European Symposium on Research in Computer Security},
  pages={589--609},
  year={2022},
  organization={Springer}
}

@article{wei2022chain,
  title={Chain-of-thought prompting elicits reasoning in large language models},
  author={Wei, Jason and Wang, Xuezhi and Schuurmans, Dale and Bosma, Maarten and Xia, Fei and Chi, Ed and Le, Quoc V and Zhou, Denny and others},
  journal={Advances in neural information processing systems},
  volume={35},
  pages={24824--24837},
  year={2022}
}

@article{deng2023pentestgpt,
  title={Pentestgpt: An llm-empowered automatic penetration testing tool},
  author={Deng, Gelei and Liu, Yi and Mayoral-Vilches, V{\'\i}ctor and Liu, Peng and Li, Yuekang and Xu, Yuan and Zhang, Tianwei and Liu, Yang and Pinzger, Martin and Rass, Stefan},
  journal={arXiv preprint arXiv:2308.06782},
  year={2023}
}

@article{cheng2023kairos,
  title={Kairos:: Practical Intrusion Detection and Investigation using Whole-system Provenance},
  author={Cheng, Zijun and Lv, Qiujian and Liang, Jinyuan and Wang, Yan and Sun, Degang and Pasquier, Thomas and Han, Xueyuan},
  journal={arXiv preprint arXiv:2308.05034},
  year={2023}
}

@inproceedings{rehman2024flash,
  title={FLASH: A Comprehensive Approach to Intrusion Detection via Provenance Graph Representation Learning},
  author={Rehman, Mati Ur and Ahmadi, Hadi and Hassan, Wajih Ul},
  booktitle={2024 IEEE Symposium on Security and Privacy (SP)},
  pages={139--139},
  year={2024},
  organization={IEEE Computer Society}
}

@misc{mitre_eval,
    title = {Turla (2023) Overview},
    howpublished = {\url{https://attackevals.mitre-engenuity.org/enterprise/turla/}},
    key = {mitre_eval},
    urldate = {2024-05-01}
}

@misc{purplesharp,
    title = {Purple Sharp},
    key = {purplesharp},
    howpublished = {\url{https://detectionlab.network/usage/purplesharp/}},
    urldate = {2024-05-01}
}

@inproceedings{alam2023looking,
  title={Looking beyond IoCs: Automatically extracting attack patterns from external CTI},
  author={Alam, Md Tanvirul and Bhusal, Dipkamal and Park, Youngja and Rastogi, Nidhi},
  booktitle={Proceedings of the 26th International Symposium on Research in Attacks, Intrusions and Defenses},
  pages={92--108},
  year={2023}
}

@misc{mitre_ttp,
    title = {TTP-Based Hunting},
    howpublished = {\url{https://www.mitre.org/sites/default/files/2021-11/prs-19-3892-ttp-based-hunting.pdf}},
    key = {mitre_ttp},
    urldate = {2024-05-01}
}

@misc{mitre_matrix,
    title = {ATT\&CK Matrix for Enterprise},
    howpublished = {\url{https://attack.mitre.org/}},
    key = {mitre_matrix},
    urldate = {2024-05-01}
}

@misc{caldera,
    title = {MITRE Caldera},
    key = {CALDERA},
    howpublished = {\url{https://github.com/mitre/caldera}},
    urldate = {2024-05-01}
}

@misc{metasploit,
    Author = {Online},
    Howpublished = {\url{https://www.metasploit.com/}},
    Title = {Metasploit: The world's most used penetration testing framework},
Year = 2025
}

@misc{atomicredteam,
    Author = {Online},
    Howpublished = {\url{https://atomicredteam.io/}},
    Title = {Explore Atomic Red Team},
Year = 2025
}

@article{xu2024autoattacker,
  title={AutoAttacker: A Large Language Model Guided System to Implement Automatic Cyber-attacks},
  author={Xu, Jiacen and Stokes, Jack W and McDonald, Geoff and Bai, Xuesong and Marshall, David and Wang, Siyue and Swaminathan, Adith and Li, Zhou},
  journal={arXiv preprint arXiv:2403.01038},
  year={2024}
}

@techreport{miller2018automated,
  title={Automated adversary emulation: A case for planning and acting with unknowns},
  author={Miller, Doug and Alford, Ron and Applebaum, Andy and Foster, Henry and Little, Caleb and Strom, Blake},
  year={2018},
  institution={MITRE CORP MCLEAN VA MCLEAN}
}

@inproceedings{takahashi2020aptgen,
  title={$\{$APTGen$\}$: An Approach towards Generating Practical Dataset Labelled with Targeted Attack Sequences},
  author={Takahashi, Yusuke and Shima, Shigeyoshi and Tanabe, Rui and Yoshioka, Katsunari},
  booktitle={13th USENIX Workshop on Cyber Security Experimentation and Test (CSET 20)},
  year={2020}
}

@inproceedings{choi2021probabilistic,
  title={Probabilistic attack sequence generation and execution based on mitre att\&ck for ics datasets},
  author={Choi, Seungoh and Yun, Jeong-Han and Min, Byung-Gil},
  booktitle={Cyber Security Experimentation and Test Workshop},
  pages={41--48},
  year={2021}
}

@INPROCEEDINGS{morse,
    author={Hossain, Md Nahid and Sheikhi, Sanaz and Sekar, R.},
    booktitle={IEEE Symposium on Security and Privacy (SP)},
    title={Combating Dependence Explosion in Forensic Analysis Using Alternative Tag Propagation Semantics},
    year={2020},
    volume={},
    number={},
    pages={}
}

@INPROCEEDINGS{holmes,  
author={Milajerdi, Sadegh M. and Gjomemo, Rigel and Eshete, Birhanu and Sekar, R. and Venkatakrishnan, V.N.}, 
booktitle={IEEE Symposium on Security and Privacy (SP)},   
title={HOLMES: Real-Time APT Detection through Correlation of Suspicious Information Flows},   
year={2019},  
volume={},  
number={},  
pages={},  
doi={10.1109/SP.2019.00026}}

@misc{noauthor_atomicredteam_nodate,
	title = {Atomic Red Team},
	url = {https://github.com/redcanaryco/atomic-red-team},
    key = {noauthor_atomicredteam_nodate},
    howpublished = "\url{https://github.com/redcanaryco/atomic-red-team}",
}

@inproceedings{nodoze,
  title={Nodoze: Combatting threat alert fatigue with automated provenance triage},
  author={Hassan, Wajih Ul and Guo, Shengjian and Li, Ding and Chen, Zhengzhang and Jee, Kangkook and Li, Zhichun and Bates, Adam},
  booktitle={network and distributed systems security symposium},
  year={2019}
}

@inproceedings{provdetector,
  title={You are what you do: Hunting stealthy malware via data provenance analysis.},
  author={Wang, Qi and Hassan, Wajih Ul and Li, Ding and Jee, Kangkook and Yu, Xiao and Zou, Kexuan and Rhee, Junghwan and Chen, Zhengzhang and Cheng, Wei and Gunter, Carl A and others},
  booktitle={NDSS},
  year={2020}
}

@INPROCEEDINGS{extractor,
  author={Satvat, Kiavash and Gjomemo, Rigel and Venkatakrishnan, V.N.},
  booktitle={2021 IEEE European Symposium on Security and Privacy (EuroS\&P)}, 
  title={Extractor: Extracting Attack Behavior from Threat Reports}, 
  year={2021},
  volume={},
  number={},
  pages={598-615},
  doi={10.1109/EuroSP51992.2021.00046}}

@inproceedings{ttpdrill,
  title={Ttpdrill: Automatic and accurate extraction of threat actions from unstructured text of cti sources},
  author={Husari, Ghaith and Al-Shaer, Ehab and Ahmed, Mohiuddin and Chu, Bill and Niu, Xi},
  booktitle={Proceedings of the 33rd annual computer security applications conference},
  pages={103--115},
  year={2017}
}

@inproceedings{hu2020automated,
  title={Automated penetration testing using deep reinforcement learning},
  author={Hu, Zhenguo and Beuran, Razvan and Tan, Yasuo},
  booktitle={2020 IEEE European Symposium on Security and Privacy Workshops (EuroS\&PW)},
  pages={2--10},
  year={2020},
  organization={IEEE}
}

@misc{LOLBAS,
	title = {Living Off The Land Binaries, Scripts and Libraries https://lolbas-project.github.io/.},
    key = {lolbas},
	url = {https://lolbas-project.github.io/}
}
